\documentclass[a4paper,11pt]{article}

\pdfoutput=1 

\usepackage{jcappub} 
\usepackage{comment}

\usepackage{subcaption}

\usepackage[T1]{fontenc} 

\makeatletter
\newcommand\makebig[2]{%
  \@xp\newcommand\@xp*\csname#1\endcsname{\bBigg@{#2}}%
  \@xp\newcommand\@xp*\csname#1l\endcsname{\@xp\mathopen\csname#1\endcsname}%
  \@xp\newcommand\@xp*\csname#1r\endcsname{\@xp\mathclose\csname#1\endcsname}%
}
\makeatother 

        \usepackage{amsfonts}
        \usepackage{amsmath}
	\usepackage{float}
\usepackage{array}
\usepackage{booktabs}
        \usepackage{colortbl}
        \usepackage{fancyhdr}
        \usepackage{textcomp}
\definecolor{dullpurple}{rgb}{0.431,0.188,0.534}
\definecolor{darkgreen}{rgb}{0.133,0.545,0.133}
\definecolor{dullred}{rgb}{0.706,0.208,0.192}

\usepackage{bm}
\usepackage{latexsym}
\usepackage{colortbl}
\usepackage{multirow}
\usepackage{array}
\usepackage{booktabs}
\usepackage{graphicx}
        \definecolor{dullpurple}{rgb}{0.431,0.188,0.534}
\hypersetup{urlcolor=dullpurple}
\hypersetup{citecolor=dullpurple}

\newcolumntype{Q}{>{$\displaystyle}l<{$}}
\newcolumntype{q}{>{\columncolor[gray]{0.9}$\displaystyle}l<{$}}
\newcolumntype{R}{>{$\displaystyle}r<{$}}
\newcolumntype{S}{>{$\displaystyle}c<{$}}
\newcolumntype{s}{>{\columncolor[gray]{0.9}$\displaystyle}c<{$}}
\newcolumntype{T}{>{\columncolor[gray]{0.9}}c<{}}

\newsavebox{\tableA}
\newsavebox{\tableB}

\newsavebox{\boxplot}
\newsavebox{\boxplota}

\makebig{biggg} {3.5}
\makebig{Biggg} {4.0}
\makebig{bigggg}{4.5}
\makebig{Bigggg}{5.5}

\title{\boldmath Entanglement masquerading\\ in the CMB}

\author[a]{Arsalan~Adil,}
\author[a]{Andreas~Albrecht,}
\author[a]{Rose~Baunach,}
\author[b,1]{R.~Holman,\note{Corresponding author.}}
\author[b]{Raquel~H.~Ribeiro,}
\author[b]{Benoit~J.~Richard}

\affiliation[a]{Center for Quantum Mathematics and Physics and Department of Physics and Astronomy, University of California at Davis, One Shields Ave, Davis, CA 95616, U.S.A.}
\affiliation[b]{Minerva University, 14 Mint Plaza, San Francisco, CA 94103, U.S.A.}

\emailAdd{aadil@ucdavis.edu}
\emailAdd{ajalbrecht@ucdavis.edu}
\emailAdd{baunach@ucdavis.edu}
\emailAdd{rh4a@andrew.cmu.edu}
\emailAdd{ribeiro@minerva.edu}
\emailAdd{brichard@minerva.edu}

\abstract{The simplest single-field inflation models capture all the relevant contributions to the patterns in the Cosmic Microwave Background~(CMB) observed today.
A key assumption in these models is that the quantum inflationary fluctuations that source such patterns are generated by a particular quantum  state---the Bunch--Davies~(BD) state.  While this is a well-motivated choice from a theoretical perspective, the question arises of whether current data can rule out other, also well motivated, choices of states. In particular, as we previously demonstrated in \cite{Baunach:2021yvu}, entanglement is naturally and inevitably dynamically generated during inflation given the presence of a ``rolling'' spectator scalar field---and the resulting entangled state will yield a primordial power spectrum with potentially measurable deviations compared to the canonical BD result.
For this work we developed a perturbative framework to allow a systematic exploration of constraints on (or detection of) entangled states with Planck CMB data using Monte Carlo techniques. We have found that most entangled states accessible with our framework are consistent with the data.  One would have to expand the framework to allow a greater variety of entangled states in order to saturate the Planck constraints and more systematically explore any preferences the data may have among the different possibilities.}

\begin{document}
\maketitle
\flushbottom

\section{Introduction}

\label{sec:intro}

The currently dominant paradigm for the formation of cosmic structure is the inflationary one~\cite{Guth:1980zm,Kazanas:1980tx,Linde:1981mu,Albrecht:1982wi}, where the universe undergoes a period of rapid expansion during which quantum fluctuations are stretched from micro to macro scales. These then freeze out after crossing the inflationary horizon, yielding a computable power spectrum for the anisotropies in the Cosmic Microwave Background radiation (CMB)~\cite{Guth:1982ec,Bardeen:1983qw}.

Going beyond the paradigmatic aspects of inflation to more concrete and detailed predictions involves model building. The simplest set of assumptions one can make is that inflation is driven by the slowly varying energy density of a single scalar, the \emph{inflaton}, and that its quantum fluctuations drive structure formation. 

At this point, we find ourselves facing a veritable smorgasbord of choices for inflationary model building. Inflaton dynamics is driven by its potential and whether or not it has a canonical kinetic term among other choices. However, for the most part, one aspect of inflationary model building seems to be fixed: the choice of quantum state for the quantum fluctuations of the inflaton. 

The so-called Bunch--Davies (BD) state~\cite{Bunch:1978yq} is this preferred state. The reasons given for this choice are sound: it is a state that respects the symmetries of de Sitter space, which is an approximation to the inflationary spacetime, it has ``good'' quantum mechanical behavior (e.g., it satisfies the Hadamard condition~\cite{BirrellDavies1982} i.e. that it has the appropriate short-distance singularity structure), and it is the state that maps into the flat space Poincaré invariant state at short distances. 

However, we would argue that these are insufficient reasons to forego considering other possible states. For example, the BD state is only one member of an infinite family of states, the so-called $\alpha$-vacua~\cite{Allen:1985ux,Mottola:1984ar} that are also de Sitter invariant. It is certainly true that the BD state is the only one of these that satisfies the Hadamard property and this has been used as a reason to discard the other $\alpha$-vacua. While we would agree that this might be a mortal flaw if this theory is treated as a UV complete theory of inflation, we are less swayed by this argument viewing it as an effective theory. Given that we only have access to what happened $55-60$ e-folds before the end of inflation, we have no idea if the inflaton is even the right degree of freedom to focus on at short distances, for example. Thus, the use of short-distance arguments to rule out choices of the initial quantum states seems to be a somewhat hubristic endeavor to us. 

We find it much more palatable to treat the choice of the BD state as \emph {exactly} that: a choice. If inflation starts at a finite time in the past we can choose other possible states for these fluctuations. The only real constraint is that the energy density in these fluctuations does not overwhelm that of the inflaton zero mode that is driving inflation. One way to enforce this is to have short inflation, i.e., have just enough inflation to define the quantum state of the observable universe.

In order to understand how good a choice the BD state is, we need to compare it to other possible choices of states. In particular, the comparison should be a physical one, i.e., one that inquires how well a putative state choice matches observations versus how well the BD state does. Our view is that we should scan among quantum states in the space of states, subject to the constraints that the state is consistent with the onset of inflation. We treat this in the spirit of an effective theory of states~\cite{Collins:2005cm,Collins:2006bg}. If this set of states is allowed by the theoretical framework, it should be examined. 

Actually scanning throughout the full space of states is a daunting task, so we will have to restrict our search to a more amenable subspace to carry out our calculations. Starting in~\cite{Albrecht:2014aga} as well as in refs.~\cite{Bolis:2016vas,Bolis:2019fmq} and culminating in ref.~\cite{Baunach:2021yvu}, we have investigated the effect of entangling the quantum state of various spectator fields such as scalars, or the tensor metric perturbations with those of the scalar metric perturbations. These states are the most general \emph{Gaussian} states one can consider involving these fields, so one only need consider the quadratic part of the action for the relevant fields in order to compute the power spectrum, a non-trivial simplification!

We can make the case that such entangled states are more generic than not; the Higgs field does in fact exist, and any extension of the standard model comes along with many scalar fields, of various masses. From this perspective, the unentangled BD state stands out as an anomaly. Our empirical motivation lies in the fact that, with precision cosmology, we might be able to identify hints of entanglement between the inflaton and this spectator field.

It’s worth understanding the implications of entanglement (of the type computed in ~\cite{Baunach:2021yvu} and in this work). Since we match to the Bunch--Davies state at the initial time (see section~\ref{sec:initialconditions}), we are {\em not} modifying the short distance properties of the theory. Rather we allow for entangled correlations to develop due to the gravitationally induced interactions between the scalar fluctuations and the spectator fields.

In ref.~\cite{Baunach:2021yvu}, we computed the \emph{TT} and \emph{TE} power spectra for the state entangling the scalar metric perturbation, $\zeta$, and a spectator scalar field, $\Sigma$, though only for sample values of the various parameters involved. To be able to make definitive statements about whether these states can offer a better explanation of the CMB data than the standard $\Lambda$CDM cosmology using the Bunch--Davies state, we need to apply Monte Carlo techniques. The latter allows us to estimate parameters of the model given Planck data.

We consider a free massive scalar field with a rolling zero mode as our spectator field. When all is said and done, the relevant parameters to estimate via Bayesian inference are: (i) the initial position and velocity of the zero mode and (ii) the ratio $m\slash H_{\rm dS}$, where $m$ is the mass of the scalar field and $H_{\rm dS}$ is the Hubble parameter during inflation. What we find for this model is that while the best fit values of these parameters are small enough to make the state almost indistinguishable from the BD state (at least from the point of view of the CMB power spectrum), there are values of these parameters that lead to an interesting phenomenology---which on the one hand yield likelihoods very close to that for the standard $\Lambda$CDM cosmology, but with ``bumps'' in the primordial power spectrum. This leads to a \emph{masquerading effect}: the state generating the primordial power spectrum could well be an entangled one, but Planck data do not provide strong enough evidence to unmask it. 

Certain aspects of our framework were constrained by requiring computational tractability.  We have found that these constraints limited the amount of entanglement we could consider and thus kept us for the most part in the ``masquerading'' regime. It remains an open question whether a more general treatment allowing greater degrees of entanglement could result in the data being more informative, and possibly even signalling a preference for particular entangled states. 

The influence of spectator fields during inflation has, of course, been explored in a variety of 
other contexts (e.g.~\cite{Braglia_2021,Braglia:2020fms, Achucarro:2010da, Cespedes:2012hu, Assassi:2013gxa, Colas:2021llj, Raveendran:2022dtb}). One of the differences between our work and some other 
approaches is that we only consider gravitational interactions between the two sectors 
(coupling between the curvature perturbations and the perturbations in the spectator scalar 
field), rather than imposing a direct coupling term in the potential (i.e. $V(\Phi,\Sigma) \neq 
V(\Phi) + V(\Sigma)$ ) or non-canonical kinetic terms linking the two sectors. The gravitational 
interaction terms arise naturally and become important when the spectator field zero mode is 
allowed to `roll’, as initially derived in~\cite{Baunach:2021yvu} and discussed in section~\ref{sec:theory} of this work. We also do not perform additional phenomenological or data-driven model building in the spectator sector to engineer additional features in the primordial power spectrum---in this paper’s analysis, all 
the features we see simply arise from a free massive scalar spectator with a rolling zero mode 
evolving in a quasi de Sitter background. Additionally, we also restrict our attention to quasi-single field models such that our formalism requires the spectator to be 
subdominant during inflation. Lastly, another difference between our work and most other 
approaches is that, by working in Schrodinger picture quantum field theory, we have explicitly 
focused our attention on evolution of the quantum state of the perturbations---which is a 
natural setting to explore entanglement.

In section~\ref{sec:theory}, we establish the theoretical foundations of the quantum state used in our calculations. This includes a summary of the work in ref.~\cite{Baunach:2021yvu} and extensions to it, notably a perturbative approach to systematically calculate the lowest order corrections to the standard inflationary power spectrum due to entanglement. 
In section~\ref{sec:methodology}, we address the foundation for our Monte Carlo analysis. In section~\ref{sec:analysis}, we present the results of our best-fit parameter estimation and analyze how our models fare against the standard $\Lambda$CDM model. We end our discussion in section~\ref{sec:conclusions} with concluding remarks and ideas for future work.

\section{Overview of entangled two-point correlators}

\label{sec:theory}

In this section, we review the theoretical framework derived in ref.~\cite{Baunach:2021yvu} and develop the perturbative approach utilized in our parameter estimation. For more details regarding entangled states please refer to refs.~\cite{Albrecht:2014aga,Bolis:2016vas,Bolis:2019fmq}, which form the basis for our analysis. The technical results in this section are valid for any choice of spectator scalar field---we restrict our focus to the free massive scalar field beginning in section~\ref{sec:methodology}.

         \subsection{Constructing the Hamiltonian}
         \label{sec:hamiltonian}
         
     Consider the system consisting of the inflaton $\Phi$, driving inflation, together with a spectator scalar field $\Sigma$. 
     The conjugate momenta to the fields are then $\Pi_{\Phi}$ for $\Phi$, and $\Pi_{\Sigma}$ for $\Sigma$. We assume that $\Phi$ and $\Sigma$ are directly uncoupled such that their corresponding potential can be linearly separated as
\begin{equation}
    \label{eq:V}
    V(\Phi, \Sigma) = V(\Phi) + V(\Sigma)\ . 
\end{equation}
In this sense, our focus lies in quasi-single field inflation models. 
The spacetime line element is taken as 
\begin{equation}
    \label{eq:dssquared}
    ds^2=-dt^2+a^2(t) \exp(2 \zeta(\vec{x},t)) d\vec{x}^2,
\end{equation}
and we neglect the tensor perturbations in this work.

We next decompose the various fields into background components, defined as their (time dependent) expectation values, and fluctuations around them. The Hamiltonian will be expanded in powers of the fluctuations and associated conjugate momenta. Defining $\eta$ to be conformal time, ranging from $-\infty$ to $0$, and $\phi(\eta) = \left\langle \Phi(\vec{x}, \eta) \right\rangle$, we work in the comoving gauge where  $\delta\phi = 0$ and $\zeta$ alone describes the scalar fluctuations that will imprint themselves in the CMB. This allows the interaction between $\Phi$ and $\Sigma$ to properly be encoded by that of $\zeta$ and $\Sigma$. 
We write the field $\Sigma(\vec{x}, \eta)$ as 
\begin{equation}
    \Sigma(\vec{x}, \eta) = \sigma(\eta) + \chi(\vec{x}, \eta)
\end{equation}
with $\sigma(\eta)=\left\langle \Sigma(\vec{x}, t) \right\rangle$. As a result the action and variables derived from it will only involve functions of the zero mode, $\sigma(\eta)$, and the momentum space counterparts of $\zeta$ and $\chi$, $\zeta_{\vec{k}}$ and $\chi_{\vec{k}}$, respectively, within this expansion treatment. 

Our one requirement of the zero mode of the inflaton is that its energy density drive a slow-roll phase of inflation. To ensure this, we will have to enforce conditions on the evolution of the spectator zero mode $\sigma(\eta)$ so that \emph{its} energy density does not interfere with inflation. We will see how this is realized below.

To compute the effect of changing the state from BD to a more general entangled state, we make use of the Schr\"odinger picture field theory. (For discussions of the Schr\"odinger picture see refs.~\cite{Boyanovsky:1993xf} and \cite{Freese:1984dv}.) The physics of the system is encoded in a wavefunctional of the form $\Psi\left[\zeta(\cdot), \Sigma(\cdot); \eta\right]$, corresponding to
a state in which the scalar metric fluctuations $\zeta$ are entangled with $\Sigma$. The wavefunctional is a solution to the Schr\"odinger equation:
\begin{equation}
    \label{eq:Schr}
    i\partial_{\eta} \Psi\left[\zeta(\cdot), \Sigma(\cdot); \eta\right]=\hat{H} \left[\Pi_{\zeta}, \Pi_{\Sigma}, \zeta, \Sigma;\eta\right] \Psi\left[\zeta(\cdot), \Sigma(\cdot); \eta\right]. 
\end{equation}
Note that eq.~\eqref{eq:Schr} involves $\zeta$ and its canonical conjugate momentum $\Pi_{\zeta}$ instead of $\Phi$ and $\Pi_{\Phi}$. This is consistent with our choice of gauge as above. As discussed in the introduction, we want to restrict ourselves to the space of Gaussian entangled states; consistency then dictates that we only keep terms up to quadratic order in the fluctuations and their canonical momenta. 

In order to construct the Hamiltonian describing our system, we first create the relevant action using \textit{MathGR} \cite{Wang:2013mea}, itself relying on the ADM formalism \cite{Arnowitt:1962hi}. For notational purposes, and to facilitate intuitive comparisons with standard inflationary literature where appropriate, we perform the following field redefinitions,
\begin{equation}
\label{eq:fieldredef}
 v_{\vec{k}}=
   z \  
    \zeta_{\vec{k}} \quad \textrm{and} \quad \theta_{\vec{k}} = a \chi_{\vec{k}}
\end{equation}
with $z(\eta)=\sqrt{2 M_{\rm Pl}^2\ \epsilon\ a^2(\eta)}$, with $\epsilon$ measuring deviations from pure de Sitter space and $M_{\rm Pl}$ being the Planck mass. We can then write the quadratic action as:
\begin{equation}
\label{eq:kspaceaction}
S=\int d\eta\ \int \frac{d^3 k}{(2\pi)^3}\ {\mathcal L}_k
\end{equation}
where
\begin{equation}
    \label{eq:kspacelagrangian}
    {\mathcal L}_k = \frac{1}{2} \vec{X}^{T \prime}_{\vec{k}}\ {\mathcal O}\ \vec{X}^{\prime}_{-\vec{k}}+\vec{X}^{T \prime}_{\vec{k}} \ {\mathcal M}_A\ \vec{X}_{-\vec{k}}-\frac{1}{2} \vec{X}^T_{\vec{k}}\ \Omega_k^2\ \vec{X}_{-\vec{k}}, 
\end{equation}
in which primes denote conformal time derivatives. The field variables are
\begin{equation}
    \label{eq:fieldvector}
     \vec{X}_{\vec{k}}= \begin{pmatrix}
        v_{\vec{k}}\\ 
        \theta_{\vec{k}}
    \end{pmatrix},
\end{equation}
and the matrices ${\mathcal O}$, ${\mathcal M}$, and the symmetric $\Omega_k^2$, to lowest-order in slow-roll, 
are given by:
\begin{subequations}\label{eq:actionmatrices}
\begin{align}
\label{eq:actionmatrices:1}
{\mathcal O} & = \begin{pmatrix}
	1  & \ -\tanh\alpha\\
	-\tanh \alpha &\ 1
	\end{pmatrix}
\\
\label{eq:actionmatrices:2}
        {\mathcal M}_A & = \dfrac{\mathcal{H}}{2} \biggg[ 
        \left(3-\epsilon + \frac{\eta_{\rm sl}}{2} \right) \tanh\alpha + 
        \dfrac{a^2\  \partial_{\sigma} V}{\mathcal{H}^2\sqrt{2 M_{Pl}^2\epsilon}}
        \biggg] \begin{pmatrix}
	0 &\ -1\\
	1 &\ 0
	\end{pmatrix}
\\
\label{eq:actionmatrices:3}
 \Omega_k^2 & =\begin{pmatrix}
	 k^2 - \dfrac{z''}{z} & & & \Omega^2_{k\ 12}\\
	\Omega^2_{k\ 12} & & &\Omega^2_{k \ 22} \\ 
	\end{pmatrix},
\end{align}
\end{subequations}
with
\begin{subequations}
\begin{align}
\Omega^2_{k \ 12}  &\equiv
-\tanh\alpha \Bigg[
k^2 +a^2\partial_{\sigma}^2 V/2 +\mathcal{H}^2 \left(1 + \frac{5\eta_{\rm sl}}{4} \right)
\Bigg] -
\mathcal{H}^2\Bigg(1 + \epsilon + \frac{\eta_{\rm sl}}{2} \Bigg)
\dfrac{a^2\  \partial_{\sigma} V}{\mathcal{H}^2\sqrt{2 M_{Pl}^2\epsilon}} \\
\Omega^2_{k \ 22}  &\equiv  k^2 +a^2 \partial^2_{\sigma}V
 -\dfrac{a''}{a} -2\epsilon (\epsilon -3) \mathcal{H}^2 \, \tanh^2 \alpha
 +4\epsilon \mathcal{H}^2 \tanh \alpha 
 \biggg( \dfrac{a^2\  \partial_{\sigma} V}{\mathcal{H}^2\sqrt{2 M_{Pl}^2\epsilon}} \biggg).
\end{align}
\end{subequations}
Above, we defined
\begin{equation}
 \tanh \alpha \equiv \dfrac{\sigma'}{\mathcal{H}\sqrt{2 M_P^2\epsilon }} \ ,  
 \label{eq:alpha_def}
\end{equation}
with the Hubble parameter $ {\mathcal H}$ in conformal time being 
\begin{equation}
{\mathcal H}\equiv \frac{a^{\prime}}{a}
\end{equation}
whereby the slow-roll parameter $\epsilon$ is given in conformal time by ${\mathcal H}^{\prime}= (1-\epsilon){\mathcal H}^2$ and $\eta_{sl}$ denotes the second slow roll parameter $\eta_{sl} \equiv \epsilon'/\mathcal{H} \epsilon$. The parametrization in terms of hyperbolic functions chosen here stems from the fact that, as discussed above, we are enforcing the subdominance of spectator zero mode dynamics relative to that of the inflaton field. This implies that $\sigma'\ll \sqrt{2 \mathcal{H}^2 M_P^2\epsilon }$ and, as such, the RHS of eq.~\eqref{eq:alpha_def} is bounded from -1 to 1. This will provide a somewhat more intuitive analysis of our results.

From the Lagrangian density in eq.~\eqref{eq:kspacelagrangian} and its relationship with the Hamiltonian density via
\begin{equation}
    \label{eq:kspacehamiltonian1}
    {\mathcal H}_k=\vec{\Pi}_{\vec{k}}^T \vec{X}^{\prime}_{-\vec{k}}-{\mathcal L}_k,
\end{equation}
one can determine the Hamiltonian density corresponding to our problem, namely,
\begin{equation}
    \label{eq:kspacehamiltonian2}
    {\mathcal H}_k = \frac{1}{2} \vec{\Pi}^T_{\vec{k}}\ {\mathcal O}^{-1}\  \vec{\Pi}_{-\vec{k}} + \vec{X}^T_{\vec{k}}\ {\mathcal M}_A^T {\mathcal O}^{-1}\ \vec{\Pi}_{-\vec{k}} + \frac{1}{2} \vec{X}^T_{\vec{k}} \left(\Omega_k^2 + {\mathcal M}_A^T  {\mathcal O}^{-1} {\mathcal M}_A\right)\vec{X}_{-\vec{k}},
\end{equation}
in which $\vec{\Pi}_{\pm \vec{k}}$ is the momentum operator conjugate to $\vec{X}_{\pm\vec{k}}$. 
The Hamiltonian is then given by 
\begin{equation}
    \label{eq:Hamiltonian}
    \hat{H}= \int \frac{d^3 k}{(2\pi)^3}\ {\mathcal H}_k
\end{equation}

For more details regarding the derivation of ${\mathcal H}_k$ from ${\mathcal L}_k$, please see \cite{Baunach:2021yvu}.\footnote{We note that the Lagrangian density specified in eqs.~\eqref{eq:kspacelagrangian} - \eqref{eq:actionmatrices:3} is equivalent to the fully Hermitian Lagrangian eventually arrived at in ref.~\cite{Baunach:2021yvu}, up to our field redefinitions and some integration by parts.}

\subsection{The Schr\"odinger equation for the entangled wavefunctional }

\label{sec:kernels}

As discussed in the introduction, we are restricting ourselves to Gaussian states that reflect an entanglement between $v$ and $\theta$, defined in eq.~\eqref{eq:fieldredef}. The most general wave functional that encodes this is given by:
\begin{equation}
\Psi \left[ \left\{v_{\vec{k}} \right\},  \left\{\theta_{\vec{k}} \right\}; \eta \right] = \mathcal{N}(\eta)\  \mathrm{\exp}
\left[
-\dfrac{1}{2} \int{\frac{d^3k}{(2\pi)^3}  \vec{X}_{\vec{k}} \ \mathcal{K}_{k}(\eta) \  \vec{X}_{-\vec{k}} }
\right]\ ,
\label{eq:wavefunctional}
\end{equation}
with $\mathcal{N}(\eta)$ serving to normalize the state and  $\mathcal{K}_{k}(\eta)$ being a matrix of kernels given by:
\begin{equation}
\mathcal{K}_{k}(\eta)  =\begin{pmatrix}
	 A_{k}(\eta) & \ C_{k}(\eta)\\
	C_{k}(\eta) & \ B_{k}(\eta) \\ 
	\end{pmatrix}\ .
 \label{eq:kernelmat}
\end{equation}
Here the off-diagonal entry $C_{k}(\eta)$ encodes the entanglement between the field fluctuations. 

We then solve the functional Schr\"odinger equation, eq.~(\ref{eq:Schr}), to generate the equation of motion for the kernels, which will specify the time evolution of our entangled state. This yields:
\begin{equation}
\label{eq:kernelEOM}
    i \partial_{\eta}{K}_{k} = K_{k}^{T} \mathcal{O}^{-1} K_{k} -\Omega_k^2 - \mathcal{M}_{A}^{T}\mathcal{O}^{-1}\mathcal{M}_{A} -i\left(\mathcal{M}_{A}\mathcal{O}^{-1}{K}_{k} + {K}_{k}^{T}\mathcal{O}^{-1}\mathcal{M}_{A}^{T} \right) . 
\end{equation}
Eq.~\eqref{eq:kernelEOM} can be decomposed into equations for the individual kernels, $A_{k}$, $B_{k}$, and $C_{k}$, given by:
\begin{flalign}
i\partial_{\eta} A_k  &= A_k^2 - \left(k^2 - \frac{z^{\prime\prime}}{z}\right) \nonumber \\
 &+ \bigggl\{\sinh \alpha\ A_k + \cosh\alpha\ C_k   + \frac{i}{2} \left[\left(3 - \epsilon + \frac{\eta_{\rm sl}}{2} \right){\mathcal H}\sinh\alpha + \frac{a^2\partial_{\sigma}V}{{\mathcal H}^2\sqrt{2 M_{\rm Pl}^2 \epsilon}}{\mathcal H}\cosh\alpha\right]\bigggr\}^2  \label{eq:kernelEqsA} \\
i\partial_{\eta} B_k  &= B_k^2 - \left(k^2 - \frac{a^{\prime\prime}}{a} + a^2\partial^2_{\sigma}V \right) \nonumber \\ 
&+ \bigggl\{\sinh\alpha\ B_k + \cosh\alpha\ C_k 
- \frac{i}{2} \left[\left(3 - \epsilon + \frac{\eta_{\rm sl}}{2} \right){\mathcal H}\sinh\alpha + \frac{a^2\partial_{\sigma}V}{{\mathcal H}^2\sqrt{2 M_{\rm Pl}^2 \epsilon}}{\mathcal H}\cosh\alpha\right]\bigggr\}^2 \nonumber \\
 &+ 2\epsilon {\mathcal H}^2\left[\left(\epsilon - 3\right)\tanh^2\alpha - 2\tanh\alpha\left(\frac{a^2\partial_{\sigma}V}{{\mathcal H}^2\sqrt{2 M_{\rm Pl}^2 \epsilon}}\right)\right] \label{eq:kernelEqsB} 
 \end{flalign}
 \begin{flalign}
i\partial_{\eta} C_k &= \cosh^2\alpha\ C_k\Big(A_k + B_k\Big) \nonumber\\
&+ \frac{\sinh 2\alpha}{2}\bigggl\{C_k^2 + A_k B_k \nonumber \\
&\hspace{6em}+ i \frac{\mathcal{H}}{2} \left[\left(3 - \epsilon + \frac{\eta_{\rm sl}}{2} \right) + \coth\alpha\frac{a^2\partial_{\sigma}V}{{\mathcal H}^2\sqrt{2 M_{\rm Pl}^2 \epsilon}}\right]\Big(B_k - A_k \Big) \nonumber \\
&\hspace{6em}+  \frac{\mathcal{H}^2}{4} \left[\left(3 - \epsilon + \frac{\eta_{\rm sl}}{2} \right)\tanh\alpha + \frac{a^2\partial_{\sigma}V}{{\mathcal H}^2\sqrt{2 M_{\rm Pl}^2 \epsilon}}\right]^2 \biggg\} \nonumber \\
&+ \tanh\alpha \biggg[k^2 + \frac{1}{2}a^2\partial^2_{\sigma}V + {\mathcal{H}}^2\left(1 + \frac{5}{4}\eta_{\rm sl}\right)\biggg] + 
{\mathcal{H}}^2 \left(1 + \epsilon + \frac{\eta_{\rm sl}}{2} \right)\frac{a^2\partial_{\sigma}V}{{\mathcal H}^2\sqrt{2 M_{\rm Pl}^2 \epsilon}} .  \label{eq:kernelEqsC}
\end{flalign}

Lastly, the equation of motion for the zero mode of the spectator field is given by:
\begin{equation}
    \label{eq:zeromode}
    \sigma''(\eta) + 2 \mathcal{H}\sigma'(\eta) +a^{2}(\eta)\partial_{\sigma}^{2}V(\sigma) = 0
\end{equation}

The equation for the entanglement kernel $C_k$ shows explicitly that either a non-zero spectator field velocity, encoded by $\tanh \alpha$, or a displacement from the minimum of its potential, described by non-zero values of $\partial_{\sigma} V$, can source a non-trivial evolution of this kernel---even if there is no entanglement present at the initial time. To solve these equations requires setting initial conditions for all the kernels and can only be done numerically. This was explored in more detail in ref.~\cite{Baunach:2021yvu} to which we refer the reader.

\subsection{Impact of entanglement in the power spectrum---a perturbative approach}

\label{sec:psperturb}

Numerical evaluation of the nonlinear, inhomogeneous, and coupled set of eqs.~(\ref{eq:kernelEqsA}~-~\ref{eq:kernelEqsC}) is clearly a formidable task and we make use of a perturbative approach to make the problem numerically tractable. As a side-effect, the perturbative expansion also allows us to better identify parameter degeneracies so that we can justifiably neglect varying the slow-roll parameters $\epsilon$ and $\eta_{sl}$ in our parameter estimation. Here, we focus on the perturbative framework and defer the discussion of identifying parameter degeneracies to section~\ref{sec:paramspacerestrict}.

The results in ref.~\cite{Baunach:2021yvu} show that for large enough values of the initial spectator parameters there will be deviations in the angular TT and TE power spectra between the Planck data and the predictions from the entangled state. This means that we could imagine constructing a perturbative expansion in terms of quantities that measure the deviation of the entangled state from the BD one. It should be noted that this simplification leads to a deviation from our original goal. Ideally, we would like to just ask the question of whether a given entangled state can be at least as consistent with the data as the standard $\Lambda$CDM cosmology. We have transformed that into the question of how much can our entangled state deviate from the BD one and still be consistent with the data, where consistency can be measured by how close the likelihood for our entangled state is to that of the BD state. 
 
We already know that the CMB angular power spectrum is in good agreement with a primordial spectrum generated by the BD state. Thus, from a perturbative standpoint, it makes sense to consider only deviations from the BD state that are not too extreme. Since these deviations are parametrized by the entanglement kernel $C_{k}$, we would like to develop a perturbative expansion around $C_{k}=0$. (We take the initial value of $C_{k}$ to be zero, so that entanglement is solely dynamically generated, as done in \cite{Baunach:2021yvu}.)

What should the control parameters for this expansion be? As stated in the previous subsection,  eq.~\eqref{eq:kernelEqsC} shows that the evolution of the spectator zero mode sources the evolution of $C_{k}$---either due to the velocity of the spectator field (through $\tanh \alpha$, via eq.~\eqref{eq:alpha_def}) or by a non-vanishing slope in the potential for the spectator field. A non-zero value for $C_{k}$, signifying entanglement, together with non-zero values for the position and velocity of the zero mode will then evolve the $A_{k}$ and $B_{k}$ kernels---corresponding to fluctuations in the inflaton and spectator fields, respectively---away from their BD equivalents. 

Guided by these insights, there are two relevant control parameters in  eqs.~(\ref{eq:kernelEqsA}~-~\ref{eq:kernelEqsC}) that emerge:
\begin{subequations}
\begin{align}
\lambda_1 & \equiv \tanh\alpha
\\
\lambda_2 & \equiv\dfrac{a^2\partial_\sigma V}{\mathcal{H}^2\sqrt{2 M_{\rm Pl}^2 \epsilon}} \  .
\end{align}
\end{subequations}
These are already required to be small so that the spectator does not dominate the expansion dynamics relative to the inflaton, so it is natural to expand the deviations of our entangled state from the BD one in powers of  $\lambda_1,\ \lambda_2$. (Note that the $\lambda_i$ evolve in time, so we demand that $|\lambda_i|< 1$ for all times specified by our perturbative equations, for all scales that contribute to the observable CMB sky today.) Specifically, we treat $\lambda_1,\ \lambda_2$ to be of the same formal order, and expand in a generic constant parameter $\lambda$ which encapsulates this (such that $\lambda_i \lesssim O(\lambda)$).\footnote{For example, in section~\ref{sec:paramspacerestrict} we take $\lambda = \lambda_{2,\mathrm{max}}$, where $\lambda_{2,\mathrm{max}}$ is the maximum value of $\lambda_2$ during the course of inflation (as derived in appendix~\ref{app:free_massive}).}

In this approximation scheme, we expand the kernels in powers of $\lambda$ as follows\footnote{There are no first-order corrections in $A_{k}$ and $B_{k}$ because there is nothing in the equations to source them, a point which we will elaborate on subsequently during our discussion of initial conditions.}:
   \begin{subequations}\label{eq:kernels_expansion}
\begin{align}
A_k & = A_k^{(0)} +\lambda^2 A_k^{(2)}
\\
B_k & = B_k^{(0)} +\lambda^2 B_k^{(2)}
\\
C_k & = \lambda C_k^{(1)} 
\end{align}
\end{subequations}
where for $C_{k}$ the expansion begins at first order in $\lambda$, because $C_{k} = 0$ is the standard single field limit. 

These expansions yield some considerable simplifications in our analysis.  First, consider the power spectrum. Following the methods in refs.~\cite{Albrecht:2014aga,Bolis:2016vas,Bolis:2019fmq} we can write the two-point function for $v_{\vec{k}} = z \zeta_{\vec{k}}$ as:
\begin{equation}
\label{eq:v2pt}
\langle v_{\vec{k}} v_{\vec{k}^{\prime}} \rangle=(2\pi)^3 \delta^{(3)}\left(\vec{k}+\vec{k}^{\prime}\right)\left( \frac{B_{k R}}{2\left(A_{k R} B_{k R}-C_{k R}^2\right)}\right)\equiv (2\pi)^3 \delta^{(3)}\left(\vec{k}+\vec{k}^{\prime}\right) P_{v}(k) \ ,
\end{equation}
which is related to the standard dimensionless inflationary power spectrum of curvature perturbations \cite{Baumann2011} via:
\begin{equation}
\label{eq:dimlessPS}
\Delta^2_s=\frac{k^3}{2\pi^2} P_{\zeta}(k) = \frac{k^3}{2\pi^2} \frac{1}{z^2} P_{v}(k) \ .
\end{equation}
Expanding $P_{v}$ to lowest order in $\lambda$, given eqs.~\eqref{eq:kernels_expansion}, we find:
\begin{equation}
    P_{v}= \frac{B_{k R}}{2\left(A_{k R} B_{k R}-C_{k R}^2\right)} = \frac{1}{2 A_{k R}^{(0)}} \left[1 + \lambda^{2} \left(\frac{-A_{k R}^{(2)}}{A_{k R}^{(0)}} +\frac{(C_{k R}^{(1)})^2}{A_{k R}^{(0)} B_{k R}^{(0)}} \right) \right] 
    \label{eq:Pv}
\end{equation}
where in the absence of entanglement $ P_v = P_{v,BD} = \frac{1}{2 A_{k R}^{(0)}}$. (The subscript `R' denotes taking the real part of the kernel.) Since the quantity we wish to compare with the CMB is ultimately $\Delta^2_s \sim P_{v}$, this tells us that neither $B_k^{(2)}$ nor higher order terms in $C_{k}$ are required for our analysis.
 
It is important to note that unlike the situation with the BD state, we are not guaranteed that $\zeta$ will remain constant outside the horizon, so the power spectrum will be evaluated explicitly at late times.

\subsubsection{Dimensionless kernel equations}

\label{sec:kerneldimless}

To proceed with the numerical analysis, we construct dimensionless versions of our equations. We begin by defining the following for dimensionless conformal time and wavenumbers:
\begin{equation}
\label{eq:dimlesstimewave}
\tau = -\frac{\eta}{\eta_0},\ q= \frac{k}{k_0} = -(1-\epsilon) k\eta_0
\end{equation}
where $\eta_0$ corresponds to the time at which the entangled evolution begins. Note that since the conformal times are all negative, $\tau$ runs from $-1$ to $0$.  We also define dimensionless kernels according to
\begin{equation}
\label{eq:dimlessKernels}
A_k(\eta) = \frac{{A}_q(\tau) }{\left(-\eta_0\right)} , \ \  B_k(\eta) = \frac{{B}_q(\tau) }{\left(-\eta_0\right)} ,\  \ C_k(\eta) = \frac{{C}_q(\tau) }{\left(-\eta_0\right)}  
\end{equation}
since they were all mass dimension 1 originally. Additionally, we define the dimensionless conformal Hubble parameter as
\begin{equation}
    \label{eq:dimlessH}
    \mathcal{H}(\eta) = \frac{\mathcal{H}(\tau) }{\left(-\eta_0\right)} , \ \  \mathcal{H}(\tau) = \frac{-1}{(1-\epsilon)\tau} \  \ 
\end{equation}
and rescale the spectator background quantities according to
\begin{equation}
\label{eq:dimlessSpec}
\sigma = s M_{pl} ,\ \  V(\sigma) = \Lambda^{4} V(s) , \ \ \mu^{2} = \frac{\Lambda^4}{H_{ds}^{2} M_{pl}^{2}} \ .
\end{equation}
as was done in ref.~\cite{Baunach:2021yvu}.

With these definitions, we construct dimensionless versions of eqs.~\eqref{eq:kernelEqsA}~--~\eqref{eq:kernelEqsC}, and then expand each equation to the lowest order in $\lambda$ required to compute eq.~\eqref{eq:Pv}.  The resulting equations are
\begin{subequations}\label{eq:dimless_kern_expand}
\begin{align}
\label{eq:dimlessA0}
i \partial_{\tau}A_{q}^{(0)} & = (A_q^{(0)})^2 - \left[ \left(\frac{q}{1-\epsilon}\right)^2 - \frac{\left(\nu_{f}^2 -\frac{1}{4}\right)}{\tau^2} \right]  
\\
i \partial_{\tau}A_{q}^{(2)} & = 2 A_q^{(0)}A_q^{(2)} \notag \\
& {} {} \quad + \left[ \tilde{\lambda}_1 A_q^{(0)} + C_q^{(1)} -\frac{i}{2(1-\epsilon)\tau} \left[ \left(3-\epsilon+\frac{\eta_{sl}}{2}\right) \tilde{\lambda}_1 + \tilde{\lambda}_2 \right] \right]^{2} \label{eq:dimlessA2}
\\
\label{eq:dimlessB0}
i \partial_{\tau}B_{q}^{(0)} & = (B_q^{(0)})^2 - \left[ \left(\frac{q}{1-\epsilon}\right)^2 - \frac{\left(\nu_{g}^2 -\frac{1}{4}\right)}{\tau^2} \right] 
\\
i \partial_{\tau}C_{q}^{(1)} & = C_{q}^{(1)} \left(A_{q}^{(0)} + B_{q}^{(0)} \right) + \tilde{\lambda}_1 A_{q}^{(0)}B_{q}^{(0)} \notag \\
& {} {} \quad + \frac{i}{2(1-\epsilon)\tau} \left[ \left(3-\epsilon+\frac{\eta_{sl}}{2}\right) \tilde{\lambda}_1 + \tilde{\lambda}_2 \right] (A_{q}^{(0)} - B_{q}^{(0)} ) \notag \\
& {} {} \quad + \tilde{\lambda}_1 \left[ \left(\frac{q}{1-\epsilon}\right)^2 +\frac{\mu^{2} \partial_{s}^{2}V(s)}{2 (1-\epsilon)^{2} \tau^{2}} +\frac{1 +\frac{5}{4}\eta_{sl}}{(1-\epsilon)^{2}\tau^{2}} \right] + \tilde{\lambda}_2 \left[\frac{1 + \epsilon + \frac{\eta_{sl}}{2}}{(1-\epsilon)^{2}\tau^{2}} \right]  \label{eq:dimlessC1}
\end{align}
\end{subequations}
where $\tilde{\lambda}_{1,2} = \frac{\lambda_{1,2}}{\lambda}$ is an algebraic simplification, and $\lambda_{1,2}$ are defined in terms of dimensionless quantities as:
\begin{subequations} \label{eq:dimlessLambdas}
\begin{align}
\lambda_1 & = \frac{(1-\epsilon)}{\sqrt{2 \epsilon}} (-\tau) \partial_{\tau}s
\\
\lambda_2 & =  \frac{\mu^{2}}{\sqrt{2 \epsilon}} \ \partial_{s}V(s) \ .
\end{align}
\end{subequations}
The solutions to the zeroth order equations for $A_{q}^{(0)}$ and $B_{q}^{(0)}$ are known;  these are just the Schr\"odinger picture version of the Mukhanov--Sasaki mode equations for the inflaton and spectator. Following~\cite{Baunach:2021yvu} we take the Hankel function indices for the BD modes of the inflaton and spectator to be:
\begin{subequations}
\begin{align}
\label{eq:Hankelnuf}
    \nu_{f} & = \frac{3}{2} + \epsilon + \frac{\eta_{sl}}{2} \\
    \nu_{g} & = \sqrt{\frac{9}{4} + 3\epsilon - \frac{\mu^{2}\partial_{s}^{2} \left . V(s)\right|_{s = s_0}}{(1-\epsilon)^{2}}} \ , \label{eq:Hankelnug}
\end{align}
\end{subequations}
with the dimensionless modes themselves and their corresponding zeroth order kernels given by:
\begin{subequations}
\label{eq:dimlessModes}
\begin{align}
    f_{v}(\tau)= & \frac{\sqrt{-\pi \tau}}{2} H_{\nu_{f}}^{(2)}\left(\frac{-q \tau}{(1-\epsilon)}\right) , \; \quad A_{q}^{(0)}(\tau) = -i\frac{f_{v}'(\tau)}{f_{v}(\tau)}\\
    g_{\theta}(\tau)= & \frac{\sqrt{-\pi \tau}}{2} H_{\nu_{g}}^{(2)}\left(\frac{-q \tau}{(1-\epsilon)}\right) , \; \quad B_{q}^{(0)}(\tau) = -i\frac{g_{\theta}'(\tau)}{g_{\theta}(\tau)} \ .
\end{align}
\end{subequations}
In contrast to the zeroth order equations, eqs.~\eqref{eq:dimlessA2} and \eqref{eq:dimlessC1} must be solved numerically, except in very special cases (see appendix~\ref{app:analytic}).

Finally, the dimensionless zero mode equation for the spectator is given by:
\begin{equation}
    \label{eq:dimlessBackground}
    s''(\tau) - \frac{2}{\tau(1-\epsilon)} s'(\tau) + \frac{\mu^{2} \partial_{s}V(s)}{\tau^{2}(1-\epsilon)^{2}} = 0 \ .
\end{equation}

\subsubsection{Setting up the initial conditions}

\label{sec:initialconditions}

Guided by the work in \cite{Baunach:2021yvu}, we construct our initial conditions for the kernel equations so that the corresponding modes are the standard Bunch--Davies ones at the initial time $\eta_{0}$ (corresponding to $\tau = -1$) at which entanglement begins to be dynamically generated. We do this via a Riccati transform. Given a kernel equation of the form
\begin{equation}
\label{eq:Ricatti}
i K^{\prime}(\tau)=\alpha_2(\tau) K^2(\tau)+\alpha_1(\tau) K(\tau)+\alpha_0(\tau).
\end{equation}
one can transform it into a mode equation of the form
\begin{equation}
\label{eq:RicattiToModeFinal}
f''(\tau)+\Omega^2 f(\tau)=0,\ \ \textrm{with} \ \  \Omega^2=\frac{1}{4}\alpha_1^2-\alpha_0 \alpha_2-\frac{i}{2} \alpha_1^{\prime}+\frac{i \alpha_1 \alpha_2^{\prime}}{2 \alpha_2}-\frac{3}{4} \left(\frac{\alpha_2^{\prime}}{\alpha_2}\right)^2+\frac{\alpha_2^{\prime \prime}}{2 \alpha_2}.
\end{equation}
by choosing
\begin{equation}
\label{eq:Ricatti_trans}
i K(\tau) = \frac{1}{\alpha_2(\tau)}\left(\frac{f'(\tau)}{f(\tau)}-\Delta(\tau)\right),
\end{equation}
with $2\Delta=i\alpha_1-\alpha_2^{\prime}\slash \alpha_2$. We thus use eq.~\eqref{eq:Ricatti_trans} to set the initial conditions for the real and imaginary parts of the kernel equations, respectively given by \cite{Baunach:2021yvu}:
\begin{subequations}\label{eq:realimKernelsIC}
\begin{align}
\label{eq:realinKernelsIC:real}
K_R(\tau_0) =& \frac{1}{2 \alpha_2(\tau_0)}\left(\frac{1}{\left| f(\tau_0)\right|^2}-\alpha_{1 R}(\tau_0)\right)\\
\label{eq:realinKernelsIC:imag}
K_I(\tau_0) =& -\frac{1}{2 \alpha_2(\tau_0)}\left(\left . \partial_{\tau}\ln\left | f\right |^2\right |_{\tau=\tau_0}+\alpha_{1 I}(\tau_0)+\left . \frac{\alpha_2^{\prime}}{\alpha_2}\right |_{\tau=\tau_0}\right) \ .
\end{align}
\end{subequations}
We need only use eqs.~\eqref{eq:realimKernelsIC} to set the initial conditions for $A_{q}^{(0)}$, $B_{q}^{(0)}$ and $A_{q}^{(2)}$---the initial condition for $C_{q}^{(1)}$ is $C_{q}^{(1)} (\tau_0 = -1) = 0$. For $A_{q}^{(0)}$ and $B_{q}^{(0)}$ the results are straightforward to compute, and we obtain:
\begin{subequations}\label{eq:realimA0B0}
\begin{align}
\label{eq:A0real}
A_{q R}^{(0)}(\tau_0 = -1) =& \frac{2}{ \pi \left |H^{(2)}_{\nu_{f}}(\frac{q}{1-\epsilon})\right|^2} \\
\label{eq:A0im}
A_{q I}^{(0)}(\tau_0 = -1) =& \frac{1}{2} \left[ 1 - 2 \nu_{f} + x \left . \left[ \frac{H^{(1)}_{\nu_{f} -1}(x)}{H^{(1)}_{\nu_{f}}(x)} + \frac{H^{(2)}_{\nu_{f}-1} (x)}{H^{(2)}_{\nu_{f}}(x)} \right] \right|_ {x=\frac{q}{(1-\epsilon)}} \right] \\
\label{eq:B0real}
B_{q R}^{(0)}(\tau_0 = -1) =& \frac{2}{ \pi \left |H^{(2)}_{\nu_{g}}(\frac{q}{1-\epsilon})\right|^2} \\
\label{eq:B0im}
B_{q I}^{(0)}(\tau_0 = -1) =& \frac{1}{2} \left[ 1 - 2 \nu_{g} + x \left . \left[ \frac{H^{(1)}_{\nu_{g} -1}(x)}{H^{(1)}_{\nu_{g}}(x)} + \frac{H^{(2)}_{\nu_{g}-1} (x)}{H^{(2)}_{\nu_{g}}(x)} \right] \right |_ {x=\frac{q}{(1-\epsilon)}} \right]  \ .
\end{align}
\end{subequations}
For $A_{q}^{(2)}$ the situation is slightly more obtuse. The easiest thing to do is to start with the dimensionless, but unexpanded, $A_{q}$ equation, i.e.,
\begin{align}
\label{eq:Aqeqn}
i \partial_{\tau} A_{q} & = A_{q}^{2} - \left[ \left(\frac{q}{1-\epsilon}\right)^2 - \frac{\left(\nu_{f}^2 -\frac{1}{4}\right)}{\tau^2} \right]  \notag \\
& {} {} \quad \quad + \frac{1}{1 - \lambda_{1}^{2}} \left[ \lambda_1 A_{q} + C_{q} -\frac{i}{2(1-\epsilon)\tau} \left[ \left( 3 - \epsilon +\frac{\eta_{sl}}{2} \right)\lambda_1 + \lambda_2 \right] \right]^{2},
\end{align}
identify the $\alpha$ coefficients (of the type in eq.~\eqref{eq:Ricatti}), use those along with the $\lambda$ expansions for $A_{q}$ and $C_{q}$ in eqs.~\eqref{eq:realinKernelsIC:real} and \eqref{eq:realinKernelsIC:imag}, and then collect the second order in $\lambda$ terms that remain.  Finally, after noting that $C_{q}(\tau_0 = -1) = 0$, we obtain:
\begin{subequations}\label{eq:realimA2}
\begin{align}
\label{eq:A2real}
A_{q R}^{(2)}(\tau_0 = -1) & = -\tilde{\lambda}_{1,0}^{2}A_{q R}^{(0)}(\tau_0 = -1) \\
\label{eq:A2im}
A_{q I}^{(2)}(\tau_0 = -1) & = -\tilde{\lambda}_{1,0}^{2}A_{q I}^{(0)}(\tau_0 = -1) -\frac{1}{2(1-\epsilon)} \left[\left(3 - \epsilon + \frac{\eta_{sl}}{2} \right) \tilde{\lambda}_{1,0}^{2} + \tilde{\lambda}_{1,0}\tilde{\lambda}_{2,0} \right] \notag \\
& {} {} \quad \quad +\left[ \frac{\eta_{sl}\tilde{\lambda}_{1,0}^{2}}{2(1-\epsilon)} + \tilde{\lambda}_{1,0}^{2} + \frac{2\tilde{\lambda}_{1,0}^{2}}{(1-\epsilon)} +\frac{\tilde{\lambda}_{1,0}\tilde{\lambda}_{2,0}}{(1-\epsilon)} \right]
\end{align}
\end{subequations}
where $\tilde{\lambda}_{1,0}$ denotes that $\tilde{\lambda}_{1}$ should be evaluated at $\tau_0 = -1$, and a term $\mathcal{O}(\eta_{sl}\epsilon)$ has been dropped from $A_{q I}^{(2)}$. From glancing at eqs.~\eqref{eq:realimA2}, one can see that $A_{q}^{(2)}$ will be zero initially, unless there is a non-zero initial velocity in the spectator zero mode (which causes $\tilde{\lambda}_{1,0}$ to be non-zero). 

To explain why there are no terms first order in $\lambda$ in eq.~\eqref{eq:kernels_expansion}, consider eq.~\eqref{eq:Aqeqn}.  If we add a term $\mathcal{O}(\lambda)$ to $A_{q}$ and expand (keeping $C_{q} = \lambda C_{q}^{(1)}$), the first order result will be
\begin{equation}
\label{eq:A1q}
i \partial_{\tau}A_{q}^{(1)} = 2 A_{q}^{(0)}A_{q}^{(1)},
\end{equation}
since the terms in the second line of eq.~\eqref{eq:Aqeqn} will always be $\mathcal{O}(\lambda^{2})$ at lowest order. The solution to this equation is $A_{q}^{(1)}(\tau) = \frac{D}{f_{q}^{2}(\tau)}$ where $D$ is an integration constant and $f_{q}(\tau)$ is the dimensionless BD mode function given by eq.~\eqref{eq:dimlessModes}, since $A_{q}^{(0)} = -i \frac{f'}{f}$ by definition. 

Then, however, if ones consults the initial conditions for $A_{q}^{(1)}$, using the same method to obtain them as was described for $A_{q}^{(2)}$ above, one would find the integration constant $D$ must vanish. This is because the only term that can be $\mathcal{O}(\lambda)$ in the initial conditions is proportional to $C_{q}^{(1)}$, and we have the condition that $C_{q}(\tau_0 = -1) = 0$.  The exact same procedure holds for the $B_{q}$ equation. Thus, unless one considers some initial entanglement---which is not part of this analysis---there are no first-order terms in $A_{q}$ and $B_{q}$ for the $\lambda$ expansion because there are no non-zero initial conditions to source them.

This completes our theoretical setup.  In the subsequent section we move to discuss what choices make our entangled state framework the most amenable to statistical parameter estimation, since our ultimate goal is to determine whether the CMB data truly prefers the Bunch--Davies state, or if it admits other entangled possibilities.

\vspace{-0.1cm}
\section{Technical perspectives and methodology}
\label{sec:methodology}

The technical work presented in section~\ref{sec:theory} outlines an approach to systematically calculate the lowest order deviations from the standard inflationary power spectrum due to an entangled state. The results so far are valid for any choice of spectator scalar field that one might wish to investigate. 

However, since our goal is to perform a proper parameter estimation with our entangled states, we must narrow our focus. The choices we made for this work are summarized in this section, and their results discussed in subsequent ones. We also comment on the physical origin of the oscillations in the entangled primordial power spectrum in section~\ref{sec:originOfoscillations}.

\subsection{Model parameters}
\label{sec:fmscalar}

We restrict our focus to the free massive scalar potential in the rest of this work, by considering
\vspace{-0.1cm}
\begin{equation}
    \label{eq:scalarPotential}
    V(\sigma) = \frac{1}{2} m_{\sigma}^{2} \sigma^{2} \ ,
    \vspace{-0.3cm}
\end{equation}
with dimensionless quantities:
\begin{equation}
\label{eq:scalarDimless}
V(s) = \frac{1}{2} s^{2} , \quad \ \Lambda^{4} = m_{\sigma}^{2} M_{pl}^{2} , \quad \ \mu^{2} = \frac{m_{\sigma}^{2}}{H_{ds}^{2}} \ .
\end{equation}
For this potential, the zero mode equation admits analytic solutions, which are discussed in appendix~\ref{app:free_massive}. 

As we initially investigated in ref.~\cite{Baunach:2021yvu}, even a simple free massive scalar potential admits a variety of different solutions for the primordial power spectrum.  If we rewrite terms in eqs.~\eqref{eq:dimlessPS} and \eqref{eq:Pv} to be\footnote{$\frac{-A_{k R}^{(2)}}{A_{k R}^{(0)}} = \frac{-A_{q R}^{(2)}}{A_{q R}^{(0)}}$ and similarly for the second term in eq.~\eqref{eq:dimlessPS2}, since all $k$ dependent kernels have mass dimension 1.}
\begin{equation}
\label{eq:dimlessPS2}
\Delta^2_s = \Delta^2_{s,BD} \left[1 + \lambda^{2} \left(\frac{-A_{k R}^{(2)}}{A_{k R}^{(0)}} +\frac{(C_{k R}^{(1)})^2}{A_{k R}^{(0)} B_{k R}^{(0)}} \right) \right] =\Delta^2_{s,BD} \left[1 + \lambda^{2} \left(\frac{-A_{q R}^{(2)}}{A_{q R}^{(0)}} +\frac{(C_{q R}^{(1)})^2}{A_{q R}^{(0)} B_{q R}^{(0)}} \right) \right], 
\end{equation}
we see the dimensionless term in the square brackets encapsulates all the lowest-order effects of entanglement on the power spectrum. For simplicity of notation, we denote this term $\Delta_{s, {\rm norm}}^2 \equiv \Delta^2_{s}/\Delta^2_{s,BD}$ to emphasize the correction effects relative to the standard BD power spectrum. The kernels themselves directly depend on the values of the slow roll parameters, $\epsilon$ and $\eta_{sl}$, as well as the mass, initial position, and initial velocity of the spectator field (as described by the equations in section~\ref{sec:theory}). We plot $\Delta_{s, {\rm norm}}^2$ for several example values of the spectator parameters below, to illustrate the richness of the parameter space.

\begin{figure}[!h]
\centering
\includegraphics[width=0.49\linewidth]{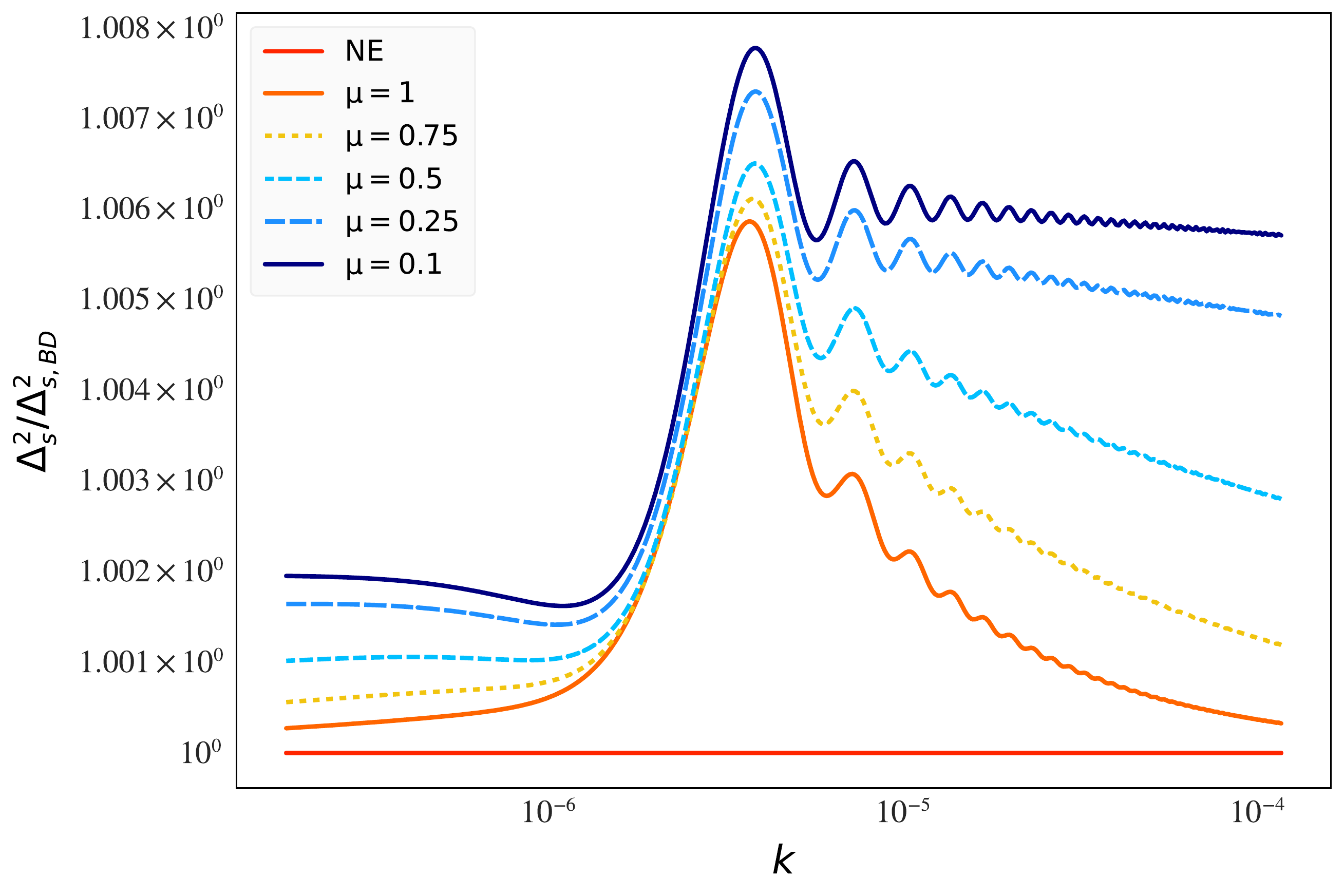} \hfill
\includegraphics[width=0.49\linewidth]{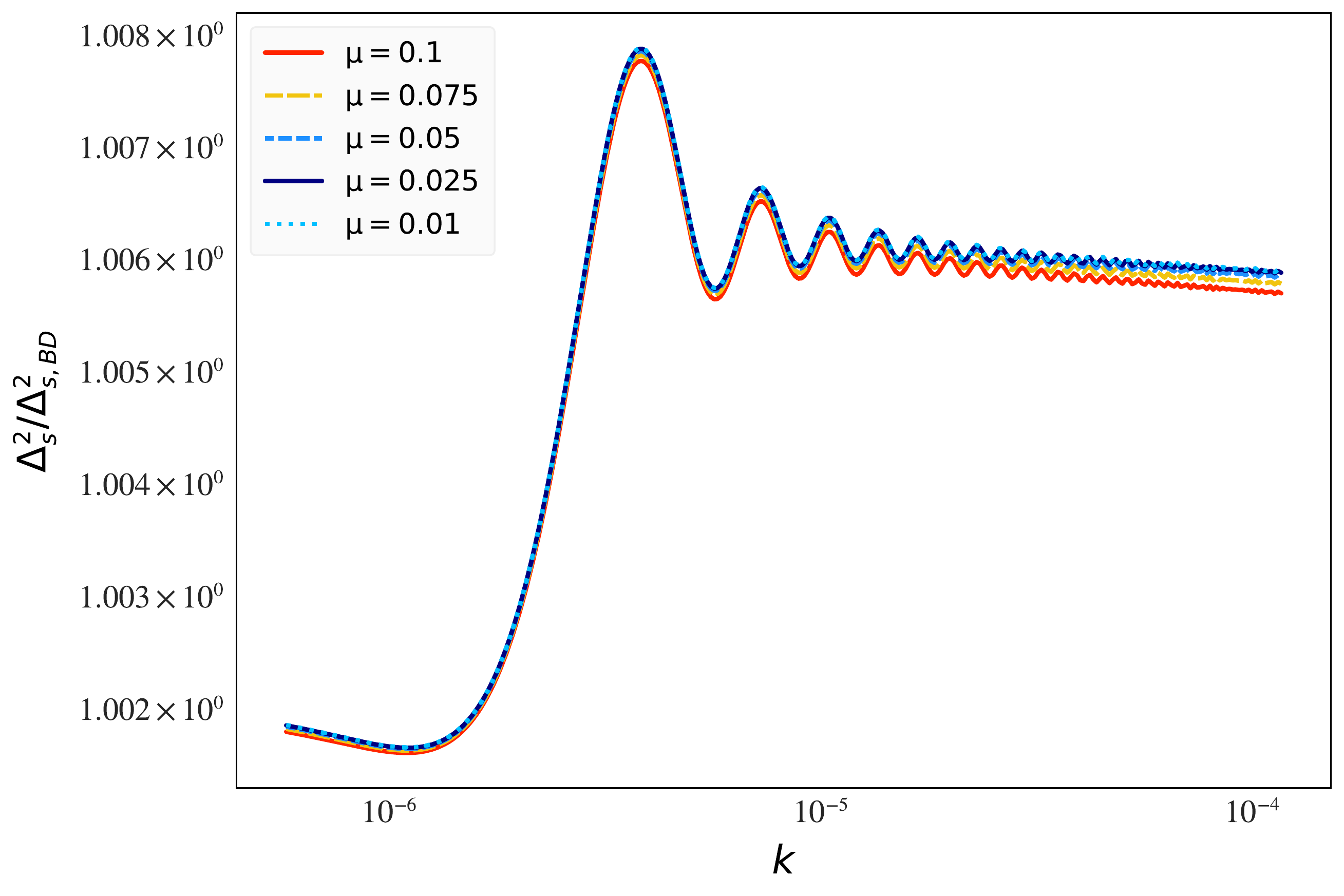}
\vspace{-0.15cm}
\caption{Log-log plots of $\Delta_{s, {\rm norm}}^2$ for a variety of dimensionless masses, given $s_0 = \frac{0.2\sqrt{2 \epsilon}}{\mu^{2}}$, $\epsilon = 10^{-7}$, and $v_0 = 0$. As discussed in the text, this choice of $s_0$ sets the expansion parameter $\lambda$ to be identical for all the curves plotted here. The non entangled (NE) case corresponds to $\Delta_{s, {\rm norm}}^2 = 1$.}
\label{fig:Muvariation}
\end{figure}

\vspace{-0.4cm}
\begin{figure}[!h]
\centering
\includegraphics[width=0.49\linewidth]{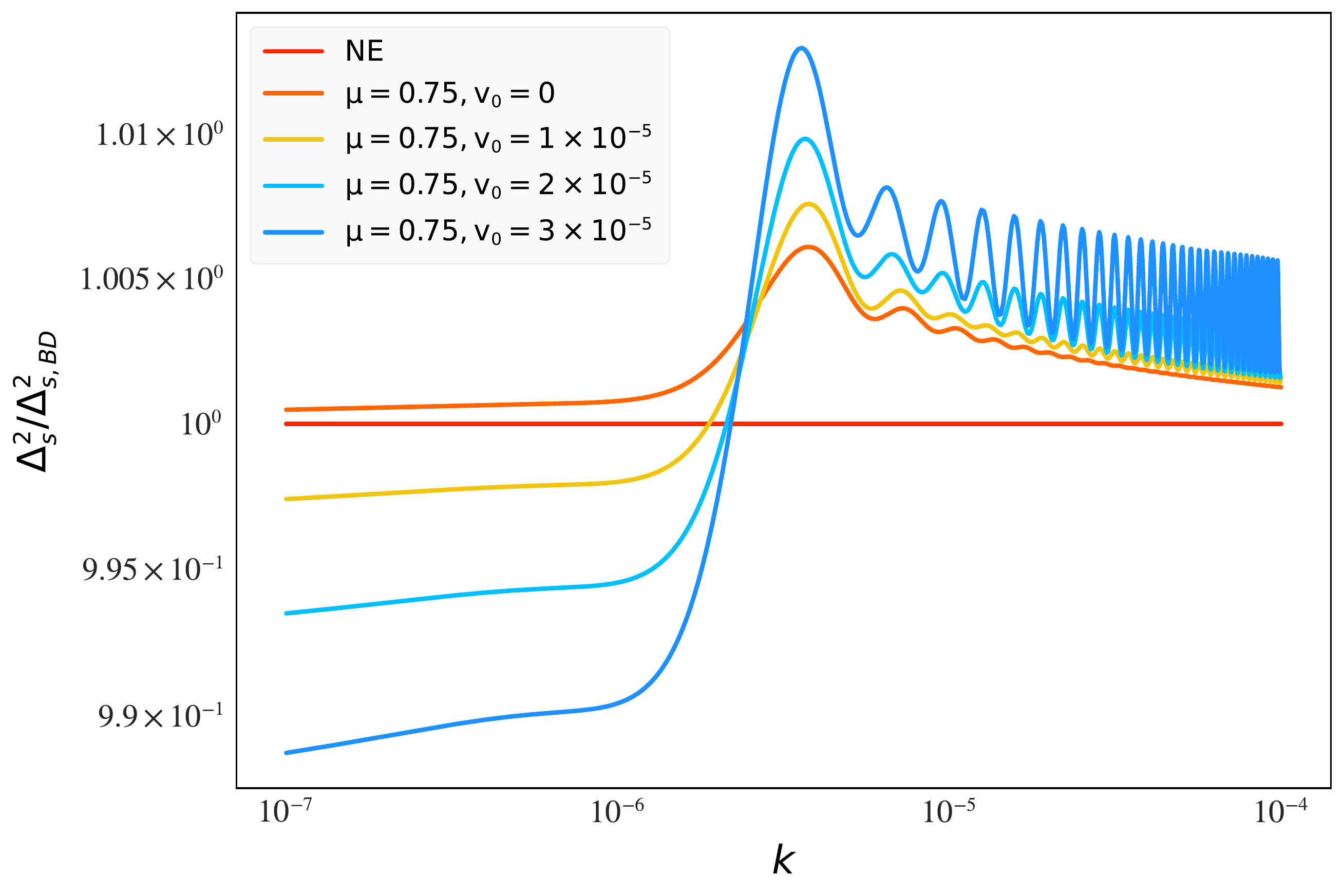} \hfill
\includegraphics[width=0.49\linewidth]{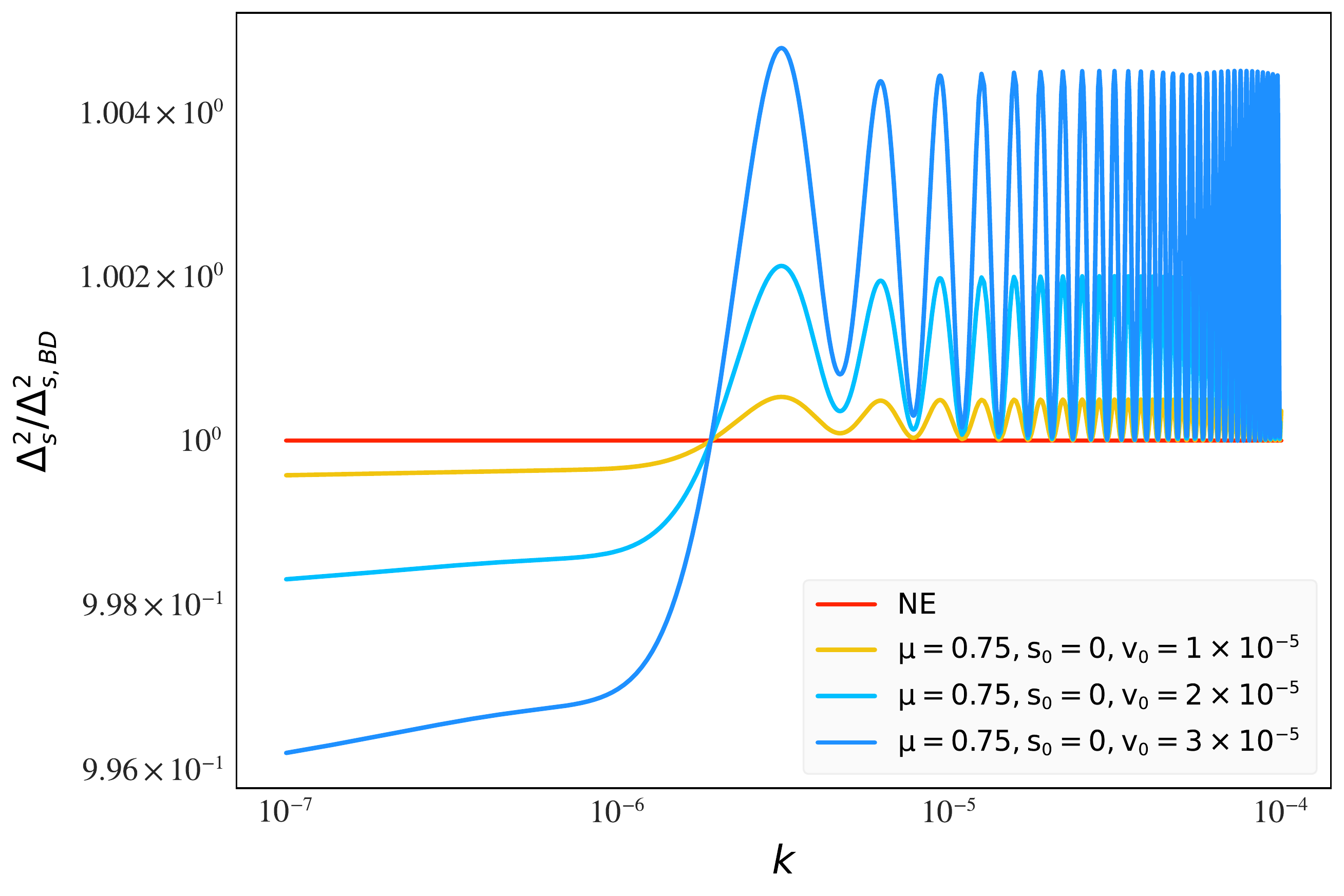}
\vspace{-0.15cm}
\caption{Log-log plots of $\Delta_{s, {\rm norm}}^2$ for $\mu = 0.75$, showing the effect of adding an initial velocity. In the plot on the left, $s_0 = \frac{0.2\sqrt{2 \epsilon}}{\mu^{2}}$ (with $\epsilon = 10^{-7}$), while $s_0 =0$ on the right. The non-entangled (NE) case corresponds to $\Delta_{s, {\rm norm}}^2 = 1$.}
\label{fig:v0variation}
\end{figure}

Figure~\ref{fig:Muvariation} shows the effect of varying $\mu$ in $\Delta_{s, {\rm norm}}^2$, given $\epsilon = 10^{-7}$ in addition to the initial conditions $s_0 = \frac{0.2\sqrt{2 \epsilon}}{\mu^{2}}$ and $v_0 = 0$. We take our expansion parameter $\lambda$ to be $\lambda = \lambda_{2,\textrm{max}} = \frac{\mu^{2}s_0}{\sqrt{2 \epsilon}}$ (where $\lambda_{2,\textrm{max}}$ is defined to be the maximum value of $\lambda_2$ during the course of inflation, given $v_0 = 0$, as discussed in appendix~\ref{app:free_massive}).  For these choices of $\lambda$ and $s_0$, one obtains $\lambda = \lambda_{2,\textrm{max}} = 0.2$ for each curve in figure~\ref{fig:Muvariation}---so that they all roughly correspond to the same amount of entanglement, isolating the effect of varying the spectator mass. As the mass of the spectator becomes lighter, differences in $\Delta_{s, {\rm norm}}^2$ get less pronounced, as shown in the right plot of figure~\ref{fig:Muvariation}.
Figure~\ref{fig:v0variation} shows the effect of adding an initial velocity, for $\mu = 0.75$, in scenarios with and without a nonzero initial position. (In figure~\ref{fig:v0variation} we take our expansion parameter to be $\lambda = \frac{\mu^{2}(s_0 + v_0)}{\sqrt{2 \epsilon}}$.)

\subsubsection{Origin of oscillations in the primordial power spectrum}
\label{sec:originOfoscillations}
In the previous section, we have explored how the mass, initial position, and initial velocity of the spectator field can influence the resulting entangled primordial power spectrum. However, one feature we have not commented on are the oscillations themselves. Oscillations in $k$ in the primordial power spectrum have been a distinctive feature of this and previous work with entangled states \cite{Baunach:2021yvu,Albrecht:2014aga,Bolis:2016vas,Bolis:2019fmq}. A glance at the equations in section~\ref{sec:theory} shows that there are several parameters that can control the placement, amplitude and persistence of these oscillations---which is also explored graphically in section~\ref{sec:fmscalar} and analytically in appendix~\ref{app:analytic}. Yet, what physically causes the oscillations in the first place? We believe we can now answer that question.

Consider what happens in standard single-field inflation. Bunch--Davies modes of different wavelengths start their evolution in phase as $\eta \rightarrow -\infty$, and continue to evolve in phase throughout their entire oscillatory regime. They reach their late time behavior after horizon crossing, upon which the modes freeze, and the standard primordial power spectrum is computed. There are no oscillations in the standard single-field result because nothing disturbs the in-phase nature of the modes throughout their entire evolution.

Let us now analyze what happens with our entangled states. We take our initial conditions to correspond to the BD modes, so before the onset of entanglement the story is the same as above---all modes begin in phase as $\eta \rightarrow -\infty$. However, at $ \eta = \eta_0$, (corresponding to $\tau_0=-1$ in dimensionless conformal time), we say that an event occurs to allow entanglement to begin \textit{for every single $k$ mode at the exact same time}.\footnote{We do not elaborate on what this special event that starts the entanglement process is. It could be a phase transition and perhaps the actual start of inflation. We chose to let our uncertainty about details of the early universe show up phenomenologically in the structure of our formalism through parameters such as $\eta_0$.} At this common initial time, every $k$ mode will be at a slightly different phase in its oscillation. This translates to an out-of-phase `initial condition' for the entangled evolution, which then propagates to late times where we compute the power spectrum. And despite all the complicated details of the modes' evolution due to entanglement between $\eta_0$ and the end of inflation, it is this assumption about the onset of entanglement that ultimately sources the $k$-dependent oscillations in the primordial power spectrum.

\begin{figure}[!h]
\centering
\includegraphics[width=0.65\linewidth]{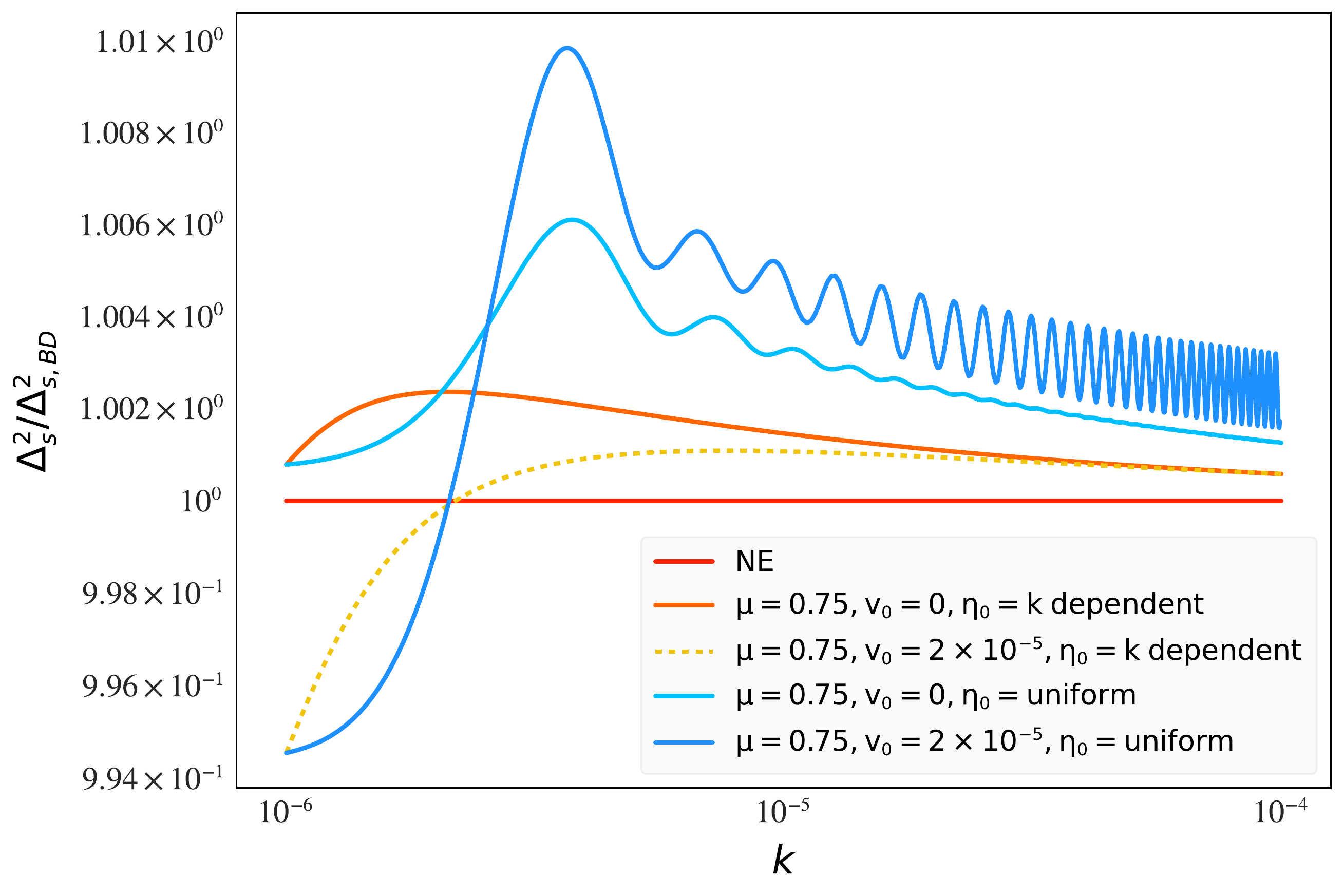}
\vspace{-0.15cm}
\caption{Log-log plot of $\Delta_{s, {\rm norm}}^2$ for $\mu = 0.75$ and $s_0 = \frac{0.2\sqrt{2 \epsilon}}{\mu^{2}}$ (taking $\epsilon = 10^{-7}$), with and without an initial velocity. We contrast the standard set-up for this paper---that all entanglement begins at the same time for each k mode---with a scenario where all the modes begin in phase, but at a different k-dependent time. As before, the non-entangled case corresponds to $\Delta_{s, {\rm norm}}^2 = 1$ and the expansion parameter $\lambda$ is the same as the one used in figure~\ref{fig:v0variation}.}
\label{fig:oscillations}
\end{figure}

We can demonstrate an explicit example of this graphically. In figure~\ref{fig:oscillations}, we plot $\Delta_{s, {\rm norm}}^2$ for $\mu = 0.75$ for two different scenarios.  The blue curves correspond to the standard set-up used throughout this paper---that the onset of entanglement happens at a fixed time for all modes.  The difference between the two is that the darker blue curve also includes an initial velocity. The orange and yellow curves instead set the onset of entanglement such that the argument of the Hankel function piece of the BD mode functions is the same for all $k$ modes\footnote{Note that the background zero mode initial condition remains the same in both scenarios since there is only one zero mode.}, effectively making $\eta_0$ $k$-dependent. The BD mode functions are given generically by
\begin{equation}
\label{eq:BDmodes}
v_{\nu}(\eta)=  \frac{\sqrt{-\pi \eta}}{2} H_{\nu}^{(2)}(-k\eta)  \; \quad v_{\nu}(\tau)=  \frac{\sqrt{-\pi \tau}}{2} H_{\nu}^{(2)}\left(\frac{-q \tau}{(1 - \epsilon)} \right)
\end{equation}
and we pick the onset of entanglement such that $x = -k \eta_0 = \frac{-q \tau_0}{(1-\epsilon)} =1$ in figure~\ref{fig:oscillations} for this scenario.

The effect of these choices is striking. For the set-up we use in this paper---that entanglement begins at a fixed time but at different phases for each of the mode functions---we see oscillations in $\Delta_{s, {\rm norm}}^2$. But, in the case where all the modes begin exactly in phase there are no oscillations in $\Delta_{s, {\rm norm}}^2$, despite the fact that $\Delta_{s, {\rm norm}}^2 \neq 1$ generically, which signifies that there is still some effect of entanglement on the primordial power spectrum. Furthermore, these conclusions hold regardless of whether the spectator has an initial velocity or not.

It's not unusual in early universe phenomenology to assume new physics happens at a given time. However, as we see from the above exploration, that assumption can have real physical consequences. If one had some physical motivation to assume that entanglement begins in a staggered way---corresponding to something like an entanglement horizon---the effects of such a scenario on cosmological observables might be quite different than what has been explored in this work.

\vspace{-0.1cm}
\subsubsection{Restricting the parameter space}
\label{sec:paramspacerestrict}

As discussed in the literature (e.g.~\cite{Braglia_2021}), choosing the best parameterization of an inflationary model to compare with data in a full parameter estimation is something of an art form. One wants to choose effective parameters---which may be the original model parameters found in one's equations  or some combination of them---whose effects on the primordial power spectrum are as distinct as possible, to preemptively eliminate as many degeneracies in the parameter estimation results as possible.

For this paper, we limit ourselves to the case where $v_0 = 0$ for the spectator field.  This choice was initially motivated by numerical difficulties evaluating highly oscillatory integrals.  However, by restricting ourselves to cases where $v_0 = 0$ and choosing $\lambda = \lambda_{2,\textrm{max}} = \frac{\mu^{2}s_0}{\sqrt{2 \epsilon}}$ as our expansion parameter\footnote{$\lambda = \lambda_{2,\textrm{max}} = \frac{\mu^{2}s_0}{\sqrt{2 \epsilon}}$ is a well-motivated expansion parameter, because one can show that $\lambda_2 < 1$ guarantees $\lambda_1 <1$ as well, for $v_0 = 0$, as discussed in appendix~\ref{app:free_massive}. }, we can effectively decouple the effects of mass and initial position on eq.~\eqref{eq:dimlessPS2}. This is because when $v_0 = 0$ and $\lambda = \frac{\mu^{2}s_0}{\sqrt{2 \epsilon}}$, $\tilde{\lambda}_{1,2} = \frac{\lambda_{1,2}}{\lambda}$ will no longer depend on either the initial position or velocity of the spectator field---as one can verify by substituting the analytical solutions in appendix~\ref{app:free_massive} into eq.~\eqref{eq:dimlessLambdas}. This eliminates all dependence on $s_0$ from the equations for $A_q^{(2)}$ and $C_q^{(1)}$, which enables us to reparameterize the scalar power spectrum as the following\footnote{Here, we have chosen the typical observational parameterization of the standard single field result, where $A_{s}$ and $n_{s}$ are the amplitude and scalar spectral index of the power spectrum, $\Delta^2_{s, BD} = A_{s} \left(\frac{k}{k_{piv}}\right)^{n_{s} - 1}$ and $k_{piv} = 0.05 Mpc^{-1}$ is the pivot scale used by Planck \cite{akrami2020planck-inflation}.}:
\begin{equation}
\label{eq:PkmuPS}
\Delta^2_s = A_{s} \left(\frac{k}{k_{piv}}\right)^{n_{s} - 1} \left[1 + \left(\frac{\mu^{2} s_{0}}{\sqrt{2 \epsilon}}\right)^{2} P_{q}(\mu, \epsilon, \eta_{sl})\right] \ 
\end{equation}
with the dimensionless quantity $P_q(\mu, \epsilon, \eta_{sl})$ defined as
\begin{equation}
\label{eq:Pkmu}
P_{q}(\mu, \epsilon, \eta_{sl}) = \left[\frac{-A_{q R}^{(2)}}{A_{q R}^{(0)}} +\frac{(C_{q R}^{(1)})^2}{A_{q R}^{(0)} B_{q R}^{(0)}} \right] \ .
\end{equation}

We next make the choices to fix $\epsilon$ and $\eta_{sl}$ to well motivated fiducial values, so that $P_{q}(\mu, \epsilon, \eta_{sl}) \equiv P_{q}(\mu)$ for our parameter estimation. Specifically, we fix:
\begin{equation}
\label{eq:eps}
\epsilon = 10^{-7}
\end{equation}
and
\begin{equation}
\label{eq:etasl}
\eta_{sl} = 1 - 2\epsilon - n_{s,f}
\end{equation}
with $n_{s,f} = 0.9649$, from the best fit values from Planck \cite{akrami2020planck-inflation}. These choices both help reduce degeneracy in our model parameter space and make our Monte Carlo analysis feasible in a timely fashion. We first comment on the degeneracy angle.

For $\epsilon < 10^{-6}$, the effect of varying $\epsilon$ in $\Delta^2_s$ is degenerate with varying $s_0$.  We have verified this numerically for a variety of values. One can also check this semi-analytically using the results in appendix~\ref{app:analytic}. Furthermore, since the tensor-to-scalar ratio---which fully determines $\epsilon$ in the slow roll expansion---is currently only an upper bound, it feels uncontroversial to fix $\epsilon$ to a fiducial value and use $s_0$ to explore our parameter space of potential spectator fields.  For the case of $\eta_{sl}$, it turns out that the effect of varying it (given $\epsilon < 10^{-6}$) in $\Delta^2_s$ can be mimicked by a combination of varying $s_0$, $\mu$, and $A_s$ for the prior ranges we consider in our analysis. We have also verified this numerically for a variety of values.

After making these choices, we decided to generate an interpolation table for $P_{q}(\mu)$ in advance of the parameter estimation run. Note that fixing $\epsilon$ and $\eta_{sl}$ to fiducial values means only 2D interpolation is required to generate the table. We find minimal integrity is lost in doing the parameter estimation this way; accuracy between the interpolated and actual numerical solutions is very high, about $10^{-5} $ percent difference at maximum. Utilizing an interpolation table also considerably speeds up the Monte Carlo analysis, as solving the equations for $A_q^{(2)}$ and $C_q^{(1)}$ is quite numerically intensive for higher wavenumbers. Also, since the entanglement parameters are uncorrelated with the standard $\Lambda$CDM ones---e.g., varying $\omega_{b}$ should have no effect on eq.~\eqref{eq:PkmuPS}---it makes little sense to recompute solutions to the primordial power spectrum for a new step in the cosmological parameter space. With the interpolation table, our parameter estimation code can avoid this, which also speeds up the analysis. 

Lastly, we include the effects of varying $\eta_0$, the onset of entanglement. As investigated in ref.~\cite{Baunach:2021yvu}, many modifications to the primordial power spectrum render the CMB observables largely unchanged when $\eta_{0}$ corresponds to the largest observable scale $k = 10^{-6} \mathrm{Mpc}^{-1}$. However, by shifting $\eta_{0}$ closer to the end of inflation, corresponding to smaller observables scales, the effects of entanglement on cosmological observables have the potential to be much more constraining.

Practically, one can see from eq.~\eqref{eq:dimlesstimewave} that shifting the initial time $\eta_{0}$ (where entanglement begins) is equivalent to shifting the scale that leaves the horizon at $\eta_0$ (the largest observable scale that will show evidence of entanglement). Therefore, the onset of entanglement can be parameterized via this distinctive scale, which we call $k_{\rm ent}$ in our analysis. And since our entangled equations are solved using dimensionless time---i.e., the dimensionless results in eq.~\eqref{eq:Pkmu} will be the same no matter what $\eta_0$ is---we can post-process our power spectrum to include the effects of shifting the onset of entanglement with a straightforward $k$ shift.  To do this, simply make the conversion
\begin{equation}
    \label{eq:kent}
    k \rightarrow k \left( \frac{k_{\rm ent}}{10^{-6} \mathrm{Mpc}^{-1}} \right) 
\end{equation}
in eq.~\eqref{eq:PkmuPS}. We take $ {k_{\rm ent}}\slash {10^{-6}} \geq 1$ in our work.

\subsection{Priors for the Monte Carlo analysis}
\label{sec:priors}

Given the choices discussed in the previous section, we are left with five parameters to vary in our parameter estimation analysis to probe the space of entangled states: $A_s$, $n_s$, $\mu$, $s_0$ and $k_{\rm ent}$. $A_s$ and $n_s$ are the standard amplitude and scalar spectral index of the primordial power spectrum, $\mu = \frac{m_{\sigma}}{H_{ds}}$ is the dimensionless mass of our spectator field, and $s_0 = \frac{\sigma_0}{M_{pl}}$ is its dimensionless initial position. Finally, $k_{\rm ent}$ is a parameter that is a proxy for adjusting the onset of entanglement, $\eta_0$, as described in the previous section.

We choose our priors for the entangled parameters as listed in table~\ref{tab:prior-range}.
	\begin{table}[h]

	\heavyrulewidth=.08em
	\lightrulewidth=.05em
	\cmidrulewidth=.03em
	\belowrulesep=.65ex
	\belowbottomsep=0pt
	\aboverulesep=.4ex
	\abovetopsep=0pt
	\cmidrulesep=\doublerulesep
	\cmidrulekern=.5em
	\defaultaddspace=.5em
	\renewcommand{\arraystretch}{1.6}

	\begin{center}
		\small
		\begin{tabular}{lll}

			\toprule
		
			 Parameter & Prior & Range \\
			\midrule
		
			\rowcolor[gray]{0.9}
				 $\textrm{log}_{10}(\mu)$ & uniform & $[-3,0]$
				\\[5mm]

				$\textrm{log}_{10}(s_{0})$ & uniform &  $[-6,2]$
				\\[5mm]

			\rowcolor[gray]{0.9}
				$\textrm{log}_{10}(k_{\rm ent})$ & uniform & $[-6,-2]$ \\

 			\bottomrule
	
		\end{tabular}
	\end{center}
 \vspace{-0.3cm}
	\caption{\label{tab:prior-range}Entanglement parameters priors and ranges.}
	\end{table}
The lower bound on $\mu$ is set so that we probe all of the interesting parameter space in figure~\ref{fig:Muvariation}, yet avoid some of the asymptotic degeneracies that set in when $\mu$ gets too small (which one can already see evidence of in the right-hand plot of figure~\ref{fig:Muvariation}). The upper bound on $\mu$ is chosen because for $\mu>1$ the Hankel function index of the spectator mode quickly becomes imaginary (see eq.~\eqref{eq:Hankelnug}), and we have been unable to find reliable numerical routines for Hankel functions of imaginary order which run fast enough to use in our Monte Carlo calculations. Unfortunately, this limitation has prevented us from exploring cases which are otherwise of interest. The value of $k_{\rm ent}$ is varied within a range that yields interesting phenomenology while being computationally feasible. All three parameters are sampled in log space since they vary over several orders of magnitude.

We additionally apply the following condition on $s_0$:
\begin{equation}
 s_0 \leq  \frac{0.5\sqrt{2 \epsilon}}{\mu^{2}},
\label{eq:s0condit}
\end{equation}
(for $\epsilon = 10^{-7}$), which acts as a joint prior on $\mu$ and $s_0$.  This condition ensures our spectator field stays subdominant to the inflaton by requiring $\lambda_{1,2} \leq 0.5$ for the full range of time our perturbative entangled evolution equations are valid. It is specifically derived from ensuring our expansion parameter $ \lambda = \lambda_{2, \mathrm{max}} \leq 0.5$ for all the values of $\mu$ and $s_0$ we investigate. This also guarantees $\lambda_{1} \leq 0.5$, for $v_0 = 0$, as discussed in appendix~\ref{app:free_massive}.

\vspace{-0.1cm} 
\subsubsection{Data and software}
\label{sec:datasoftware}

For our Monte Carlo analysis we focus on Planck data, since the CMB is the standard probe of primordial effects from inflation. We use the Planck 2018 high- and low-$\ell$ temperature and polarization likelihoods \cite{2020Planck}. For the high-$\ell$ spectra, we make use of the plik-lite code which differs from the full plik likelihood in the number of nuisance parameters.

Our software choices for our analysis are as follows. We use the Cosmic Linear Anisotropy System Solver
(CLASS) \cite{Blas_2011} as our Einstein-Boltzmann solver, and MontePython \cite{Brinckmann:2018cvx,Audren_2013} to sample the parameter space. To explore the effects of our entangled states we use the MultiNest \cite{Feroz2008,Feroz2009, Feroz_2019} sampler as implemented in MontePython via PyMultiNest \cite{Buchner2014}. We use a low evidence threshold of $10^{-5}$ in MultiNest to reliably perform the likelihood analysis in section~\ref{sec:analysis} \cite{feroz2011likelihood}.  Finally, we use GetDist \cite{Lewis:2019xzd} to generate our posterior distribution plots.

\vspace{-0.1cm} 
\section{Analysis and insights}
\label{sec:analysis}

In this section, we report our results investigating whether CMB data prefer a primordial power spectrum generated by the BD vacuum or by an entangled state.  As detailed in section~\ref{sec:methodology}, we have restricted our focus to entanglement generated due to a free massive spectator field with no initial velocity in its zero mode. Within the limits of our exploration (the scope of which is significantly bounded by numerical considerations) we have found that for the most part the effects of entanglement are too small to be constrained by (or favored by) the CMB data.  Thus, we say that the entangled states we consider are ``masquerading'' as BD states.

\subsection{Bayesian inference results}
\begin{figure}[hbtp]
\centering
\includegraphics[width=0.9\linewidth]{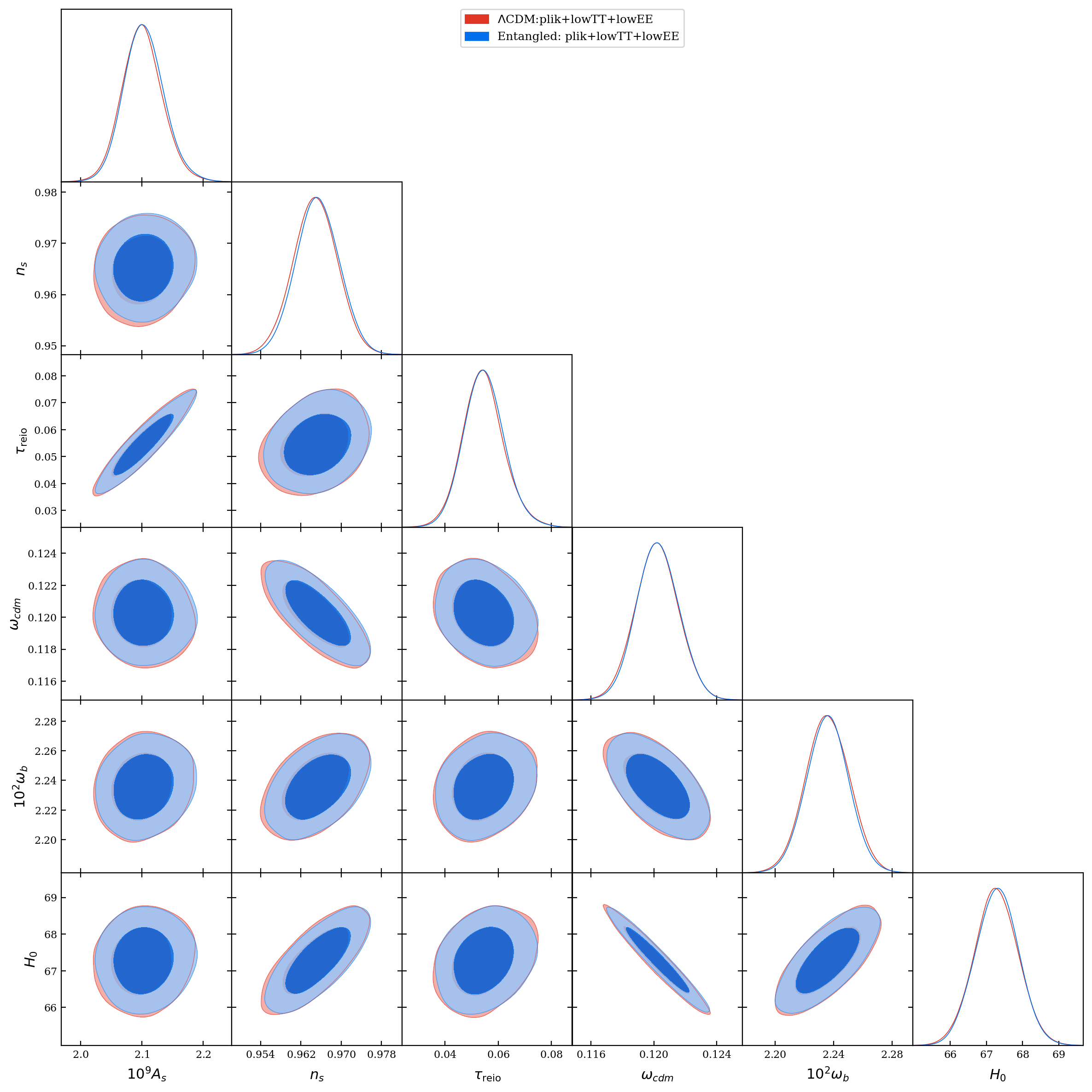}
\vspace{-0.15cm}
\caption{Posterior distributions for the cosmological parameters $A_s$, $n_s$, $\tau_{reio}$, $\omega_{b}$, $\omega_{cdm}$ and $H_{0}$. We compare the standard $\Lambda$CDM inflationary scenario, whose primordial power spectrum is generated by the BD vacuum state, versus a primordial power spectrum generated by an entangled state (parameterized by eq.~\eqref{eq:PkmuPS}).}
\label{fig:cosmo_dist}
\end{figure}

\begin{figure}[hbtp]
\centering
\includegraphics[width=0.75\linewidth]{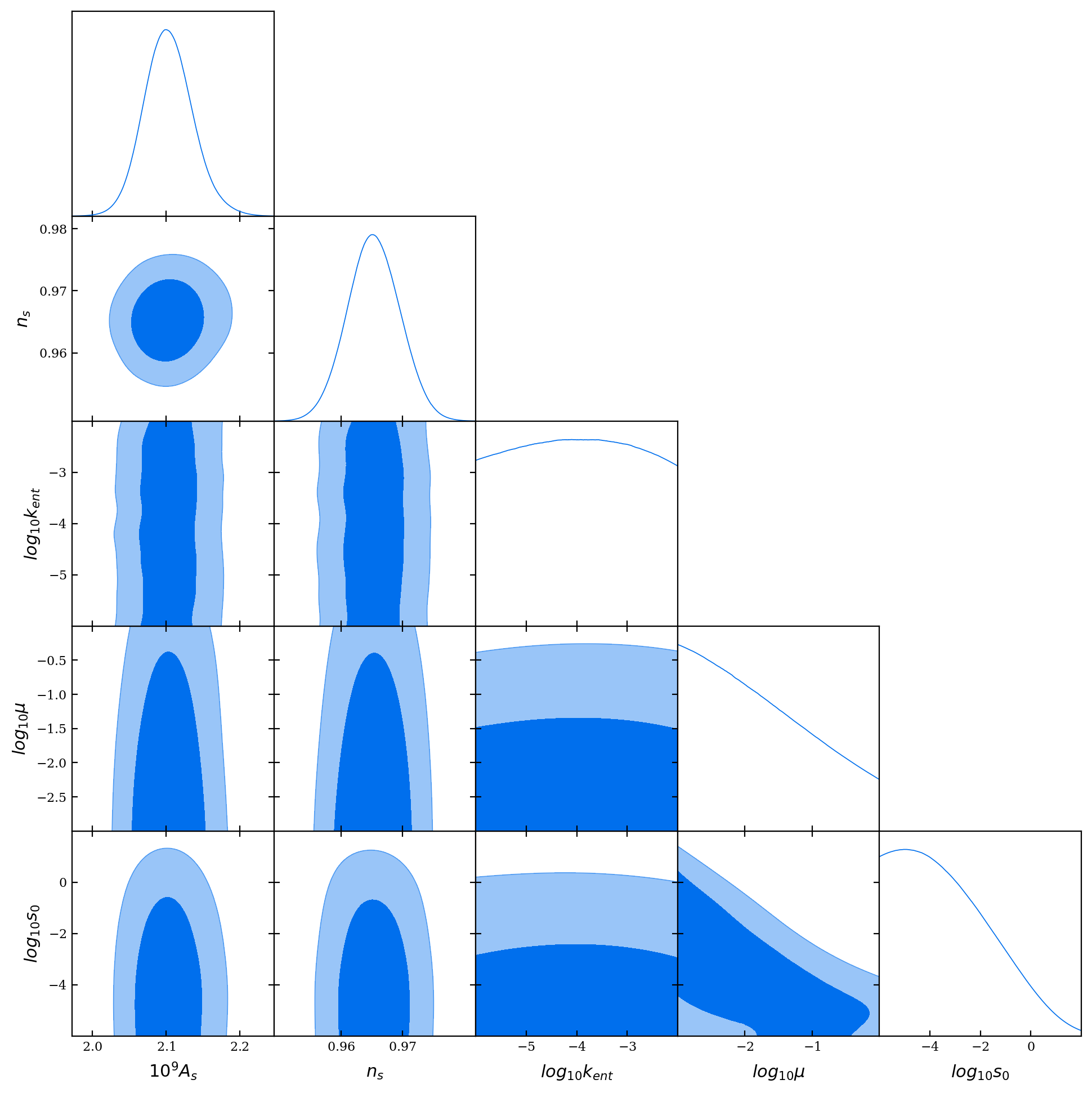}
\vspace{-0.15cm}
\caption{Posterior distributions for the entangled power spectrum parameters $A_s$, $n_s$, $\log{k_{ent}}$, $\log{\mu}$ and $\log{s_0}$, as defined in eq.~\eqref{eq:PkmuPS}. The corresponding numerical values characterizing the distributions are listed in table~\ref{tab:posteriorEtc} and $\chi^2$ values are listed in table~\ref{tab:lambdaCDMcompare}.}
\label{fig:entang_dist}
\end{figure}

Figures~\ref{fig:cosmo_dist} and \ref{fig:entang_dist} show the resulting posterior distributions in our Monte Carlo study for the usual $\Lambda$CDM cosmological parameters and those pertinent to the entangled spectrum respectively. In figure~\ref{fig:entang_dist}, one can see the posterior probability favors low values of $\mu$ and $s_0$, corresponding to the region of parameter space that is asymptotically degenerate with the BD state. However, as we explain in this section and appendix~\ref{sec:appendix-prior-volume}, this preference is largely prior driven. 

We first turn to the constraints on the cosmological parameters in figure~\ref{fig:cosmo_dist}. Clearly, there is negligible change in this distribution compared to the standard $\Lambda$CDM model, i.e., assuming a BD primordial spectrum. 
This result is straightforward to understand since,
\vspace{-0.4cm}
\begin{equation}
    C_\ell \propto \int \frac{dk}{k} \Delta_s^2(k) T_\ell^2(k, \overline{\eta_0})
    \label{eq:cls}
\end{equation}
where $T_\ell(k)$ is the transfer function which can be understood as the operator responsible for projecting the three-dimensional Fourier modes onto the two-dimensional spherical last-scattering surface and propagating the CMB photons from recombination ($\overline{\eta_*}$) to us today ($\overline{\eta_0}$). Here, we only focus on the temperature $C_\ell$ \cite{hu1997damping}. Notably, all the information of the cosmological parameters $\omega_b$, $\omega_{\rm cdm}$, $H_0$, and $\tau_{\rm reio}$ is contained in $T_\ell(k,\overline{\eta_0})$. However, one may wonder if the inference of these parameters can be ``confused'' by inducing changes in the primordial spectrum such that the cosmological parameters shift proportionately to leave the observable quantity, the $C_\ell$, unchanged. Though not impossible \cite{Hazra:2022rdl}, this seems unlikely since each of the parameters leave distinct fingerprints on the CMB spectrum (see e.g. \cite{pan2016cosmic}): $\omega_b$ affects the even-odd peak modulation through the baryon loading effect \cite{hu2002cosmic}; $\omega_{\rm cdm}$, amongst other features, determines the matter-radiation equality scale which is directly revealed by the radiation driving envelope \footnote{While we have not rigorously investigated this, one may wonder if it is possible to leave $\theta_{\rm eq}$, the comoving size of the horizon at matter-radiation equality (projected from the last-scattering surface), invariant by boosting power at $k \gtrsim 10^{-2} \approx l_{\rm eq} /\overline{\eta_0}$ accompanied by a commensurate decrease in $\omega_{\rm m}$. While interesting, this is unlikely to change the inference of $\omega_{\rm cdm}$, not only because the parameter is constrained by other effects (prominently the integrated Sachs-Wolfe and the CMB damping tail) \cite{aghanim2017planck-parameter-shift}, but also because, as pointed out in \cite{Adil:2022hkj}, the entire shape of the radiation-driving envelope, and not just $\theta_{\rm eq}$, is sensitive to changes in $\omega_{\rm m}$.}; and $H_0$ affects the spacing of the acoustic peaks via its impact on the angular size of the sound horizon at recombination. Thus, even \textit{a-priori}, judging from some sample power spectra in figure~\ref{fig:Muvariation}, it appears unlikely that the features introduced by our perturbative approach to entanglement can mimic any of the imprints induced by variations in the usual cosmological parameters. This reasoning supports the inferences seen in figure~\ref{fig:cosmo_dist}. 

Next, we turn to the constraints on the entanglement parameters in figure~\ref{fig:entang_dist}. 
First, consider the constraints (or lack thereof) on $k_{\rm ent}$. The uniform distribution on $k_{\rm ent}$ occurs because we vary $\mu$ and $s_0$ over a large range, most notably also sampling small values for these parameters which consequently lead to negligibly small deviations from the BD spectrum. Thus, if these distinguishing features are small, it does not matter where on the spectrum they occur as they would all lie in the Planck error budget.

\begin{figure}[hbtp]
\centering
\includegraphics[width=0.42\linewidth]{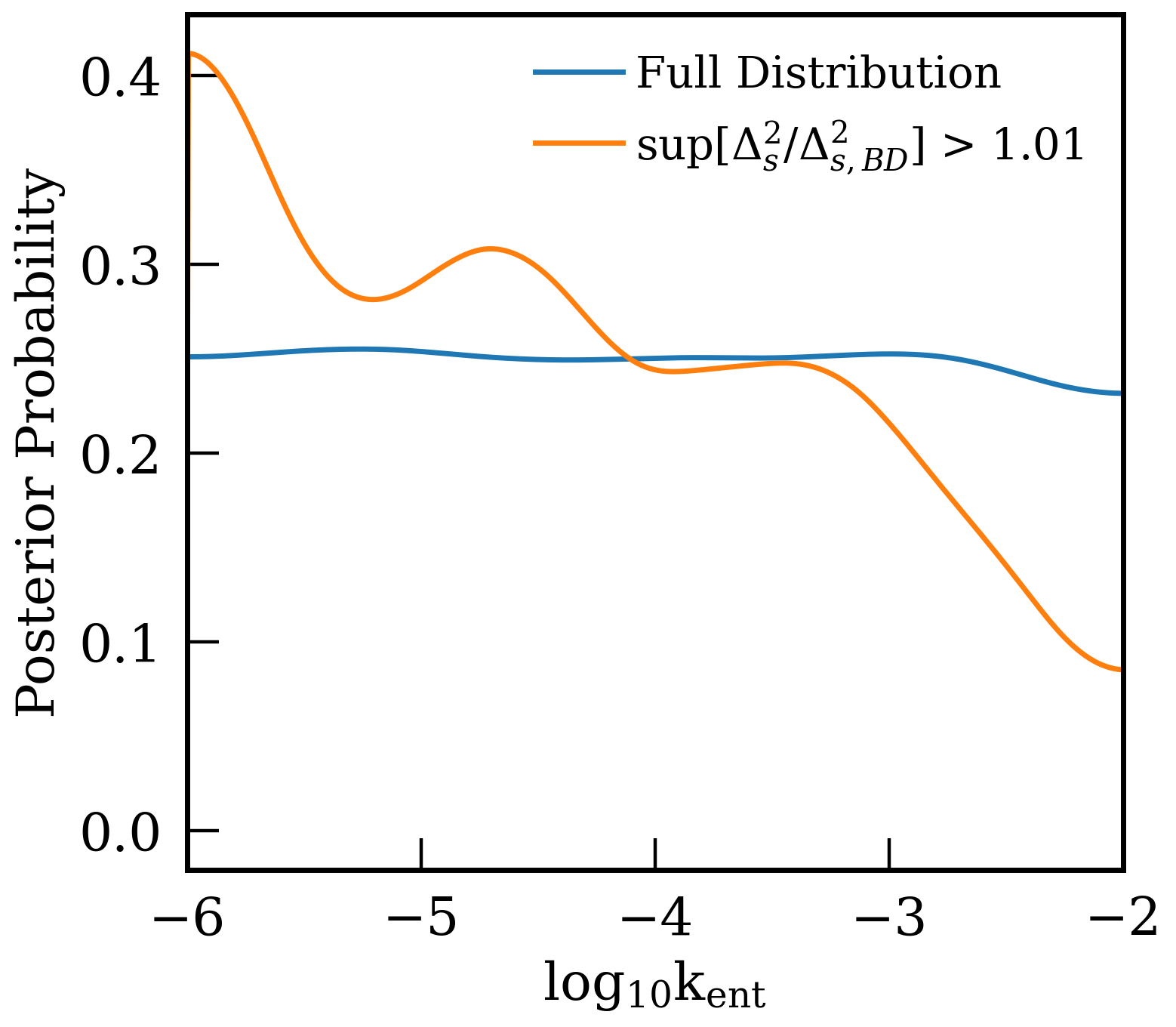} \hfill
\includegraphics[width=0.56\linewidth]{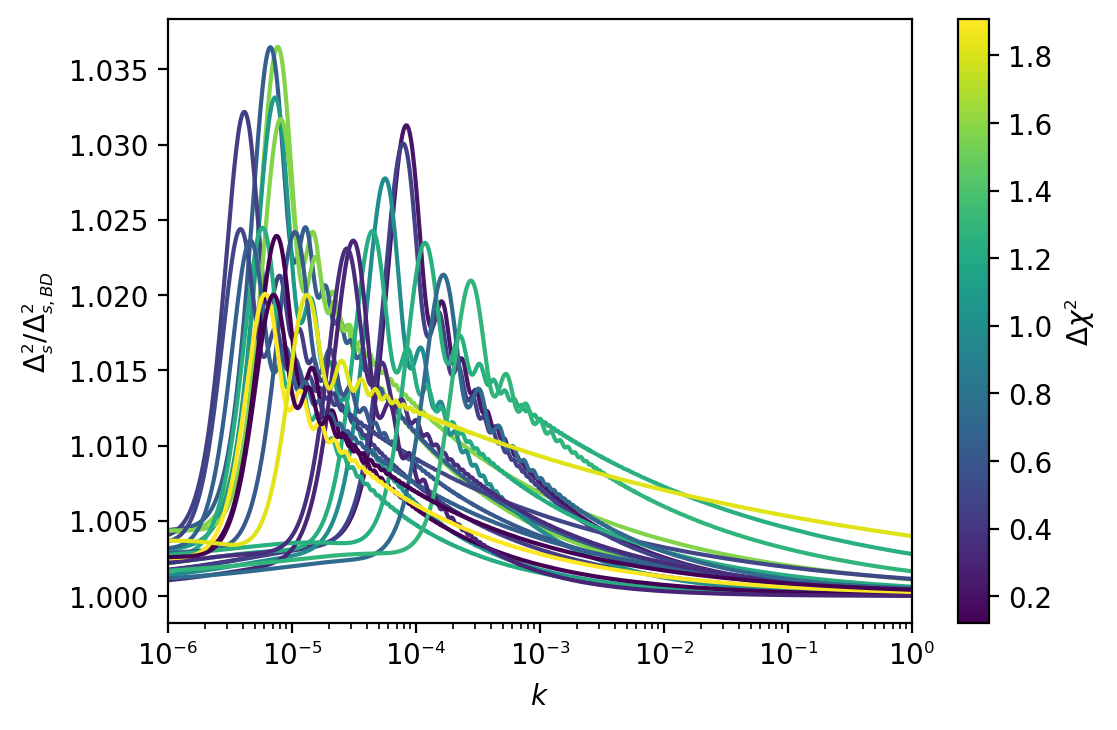}
\vspace{-0.15cm}
\caption{To investigate the nature of the spectra with the highest deviations from BD, on the right we show various entanglement primordial power spectra that deviate by at least $2\%$ from BD. Note that the samples are drawn from our Monte Carlo chain so that the $\chi^2$ has dependence on the cosmological parameters. Therefore, we also limit the sample to points with $\Delta\chi^2<2$ (relative to the best-fit) so as to minimize the impact of this dependence. The plot on the left shows the distribution in $\log{k_{\rm ent}}$ if only spectra that deviate by at least $1\%$ from BD are taken into account.}
\label{fig:k_ent_limits}
\end{figure}

But what about the larger deviations? If the deviations from BD are significant, does the data then prefer where in $k$ these features appear? We address this question by imposing a cut-off and retaining only those sets of parameters that lead to a maximum deviation of at least $2\%$ from the BD spectrum, i.e., $\sup[\Delta_{\rm s,norm}^2 > 1.02]$. The resulting power spectra are shown in the right panel of figure~\ref{fig:k_ent_limits} (note that we additionally impose a cut-off of $\Delta \chi^2 < 2$, relative to the best-fit point, to control for the effect of the variation in the other parameters) and indicate that, while non-negligible deviations of up to $\approx 3\%$ fit the Planck data about as well as the best-fit point, that these deviations preferentially occur in the $k\lesssim 10^{-3}$ regime. In the left-panel of figure~\ref{fig:k_ent_limits}, we contrast the uniform distribution on $k_{\rm ent}$ from figure~\ref{fig:entang_dist} to the distribution generated by seeking only those points that deviate from BD spectrum by at least $1\%$, which lends evidence to the suspicion garnered from the plot on its right. This skewed distribution, indicating that the highest deviations from BD occur at the large scales, likely reflects the pronounced cosmic variance at $\ell \lesssim 30$.  
Following the discussion in section~\ref{sec:paramspacerestrict}, this is essentially a constraint on when, during inflation, the most highly entangled states can emerge. The $68\%$ ($95\%$) highest posterior density interval for the reduced $k_{\mathrm{ent}}$ distribution in figure~\ref{fig:k_ent_limits} is $\mathrm{log}_{10}(k_{\mathrm{ent}}) < -3.6 \; (<-2.5)$ (given the prior range $\mathrm{log}_{10}(k_{\rm ent}) \in [-6,-2]$).

\begin{figure}[!h]
\centering
\includegraphics[width=0.53\linewidth]{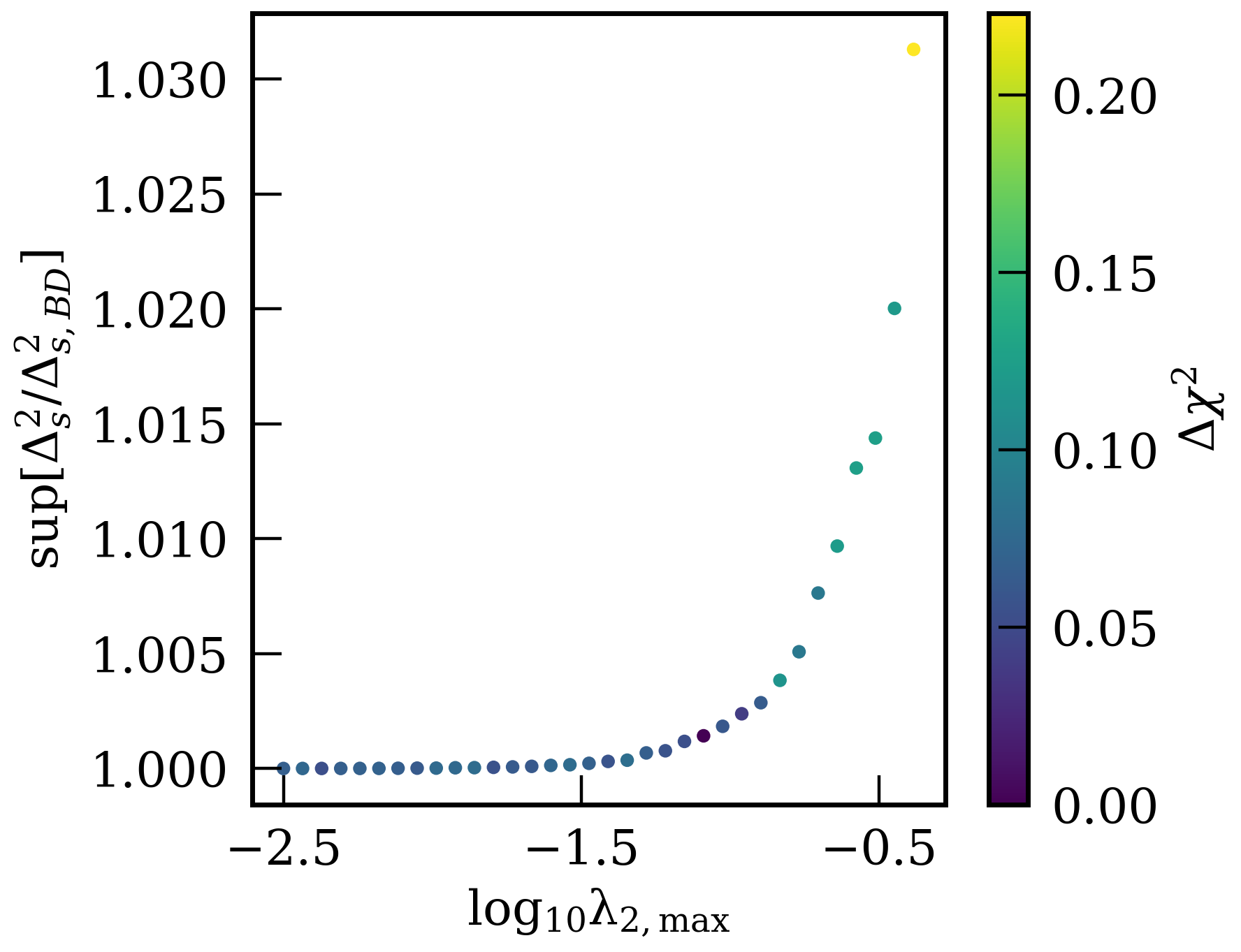}
\vspace{-0.15cm}
\caption{Here we depict the maximum deviation from the BD state as a function of $\lambda_{2,max}$. Since the samples are drawn from our Monte Carlo chain, for each bin in $\lambda_{2,max}$ we pick the point with the smallest $\chi^2$ (relative to the best-fit) so as to project out the dependence on the other parameters.}
\label{fig:lambda_scatter}
\end{figure}

Finally, we turn towards the constraints on the two parameters controlling the dynamics of the spectator field: the dimensionless mass $\mu$ and the initial position of the spectator $s_0$. The posteriors in figure~\ref{fig:entang_dist} may lead one to erroneously conclude that the data have a preference for lower $\mu$ and $s_0$. However, the profile likelihood for these parameters paints a very different picture than the posterior distribution: both are uniformly distributed in the entire range allowed by our perturbative expansion 
(see table~\ref{tab:prior-range}). From a strictly ``frequentist'' perspective, this uniform distribution makes sense: large values of $\mu$ can be compensated by small values of $s_0$ (and vice versa) so that the novel features in the primordial spectrum remain small. In fact, as shown in figure~\ref{fig:lambda_scatter}, even when there is significant deviation from BD, the $\chi^2$ remains approximately the same. This same point is also illustrated in the right panel of figure~\ref{fig:k_ent_limits}. Why then do the posteriors differ from a uniform distribution? The skewness in the posterior is due \textit{entirely} to the prior. It  stems from the condition in eq.~\eqref{eq:s0condit} which effectively acts as a joint prior on $\mu$ and $s_0$. 

\begin{figure}[hbtp]
\centering
\includegraphics[width=0.66\linewidth]{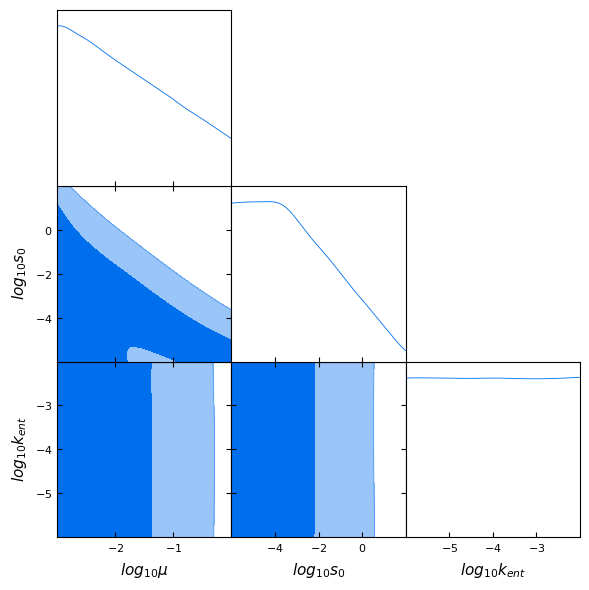}
\vspace{-0.15cm}
\caption{Posterior distributions on the entanglement parameters generated in the absence of data by assuming uniform likelihood on the entanglement parameters but imposing the same priors as those used for the posteriors in figure~\ref{fig:entang_dist}. }
\label{fig:only_prior_ent_dist}
\end{figure}

To better understand the effect of this prior, let us \textit{assume} that we are given a likelihood that is uniform in $\log\mu, \log s_0$, and, for completeness, suppose in $\log(k_{\rm ent})$ as well. We then impose the same priors on these parameters as those listed in table~\ref{tab:prior-range}, along with the condition eq.~\eqref{eq:s0condit}. The resulting posterior distributions for this toy example are shown in figure~\ref{fig:only_prior_ent_dist} and match almost exactly what we see in figure~\ref{fig:entang_dist} (note that now we do not make use of any data!). This allows us to conclude that the posterior distribution on $\log\mu$ and $\log s_0$ is entirely prior driven. In fact, the skewed distribution in $\log \mu$ owes itself, at least partially, to a prior-volume effect: because of the condition eq.~\eqref{eq:s0condit}, smaller values of $\mu$ have more prior volume in $s_0$ available, while having an equal likelihood, which assigns more posterior weight to the low-$\mu$ regime upon marginalizing over $s_0$. 

Certainly the issue of priors affecting parameter inferences is not a new one; the cosmology literature alone has many examples of this effect (e.g. on the inference of inflationary parameters, on the inference of the number of ultra-relativistic species, on dark matter interactions, and recently on the significance of early-dark energy to ameliorate the $H_0$ cosmological tension \cite{ballesteros2008flat, Gonzalez-Morales:2011tyq, heavens2018objective, Diacoumis:2018ezi}). There are several tools available to address such effects. Here we use two methods. First, we do a likelihood profile analysis (see e.g. \cite{Herold:2022iib}) and find near uniform distributions on all the entanglement parameters. The results are summarized in table~\ref{tab:posteriorEtc}. This analysis is completely decoupled from priors and strictly tells us that the probability of the Planck data being generated by a BD state is (approximately) the same as the probability of it being generated by an entangled state. Second, we run an MCMC assuming an iso-likelihood, along with our priors on the entanglement parameters, which gives identical results to the posteriors generated using the Planck data (shown in figure~\ref{fig:only_prior_ent_dist}). This reaffirms the profile likelihood analysis indicating that the data are not informative on the model parameters.

Despite the caveats in interpreting the posterior distributions at face value, they nevertheless contain important information of the underlying physics. On a physical level, the posterior of $\mu$ being driven to lower values incorporates a penalty for fine-tuning: higher $\mu$ values require the initial condition of the spectator field to lie in a smaller phase-space volume compared to lower values of $\mu$. Very importantly, our priors on both $\mu$ and $s_0$ are guided in part by fundamental physics constraints  
such that the spectator field remain subdominant to the inflaton energy. These \textit{physical} constraints lead to the joint prior on $\mu- s_0$ (eq.~\eqref{eq:s0condit}) (see appendix~\ref{sec:appendix-prior-volume} for the effect of increasing the prior volume in $s_0$ on the parameter inferences). This is in contrast to the oft-studied parameterized/reconstruction approach to the primordial power spectrum (e.g. \cite{bridle2003reconstructing, mukherjee2005reconstruct, akrami2020planck-inflation}) where the significance of a particular parameterization, and therefore that of the parameters on which the prior is enforced, is arbitrary and, consequently, the interpretation of prior-dominated posteriors is equally arbitrary \cite{trotta2008bayes}. 

\vspace{-0.2cm}

\begin{figure}[hbtp]
\centering
\includegraphics[width=0.62\linewidth]{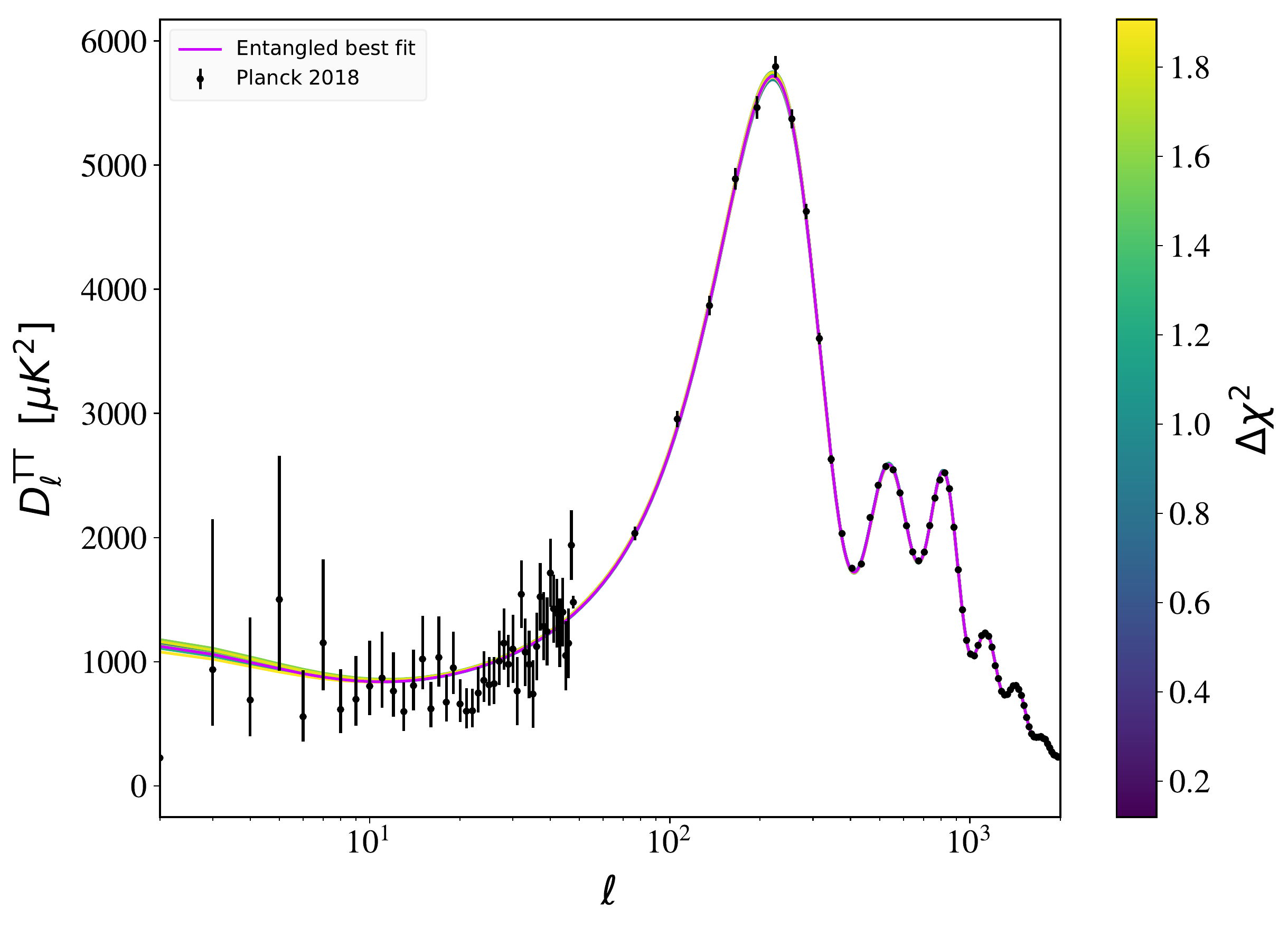} \hfill
\includegraphics[width=0.62\linewidth]{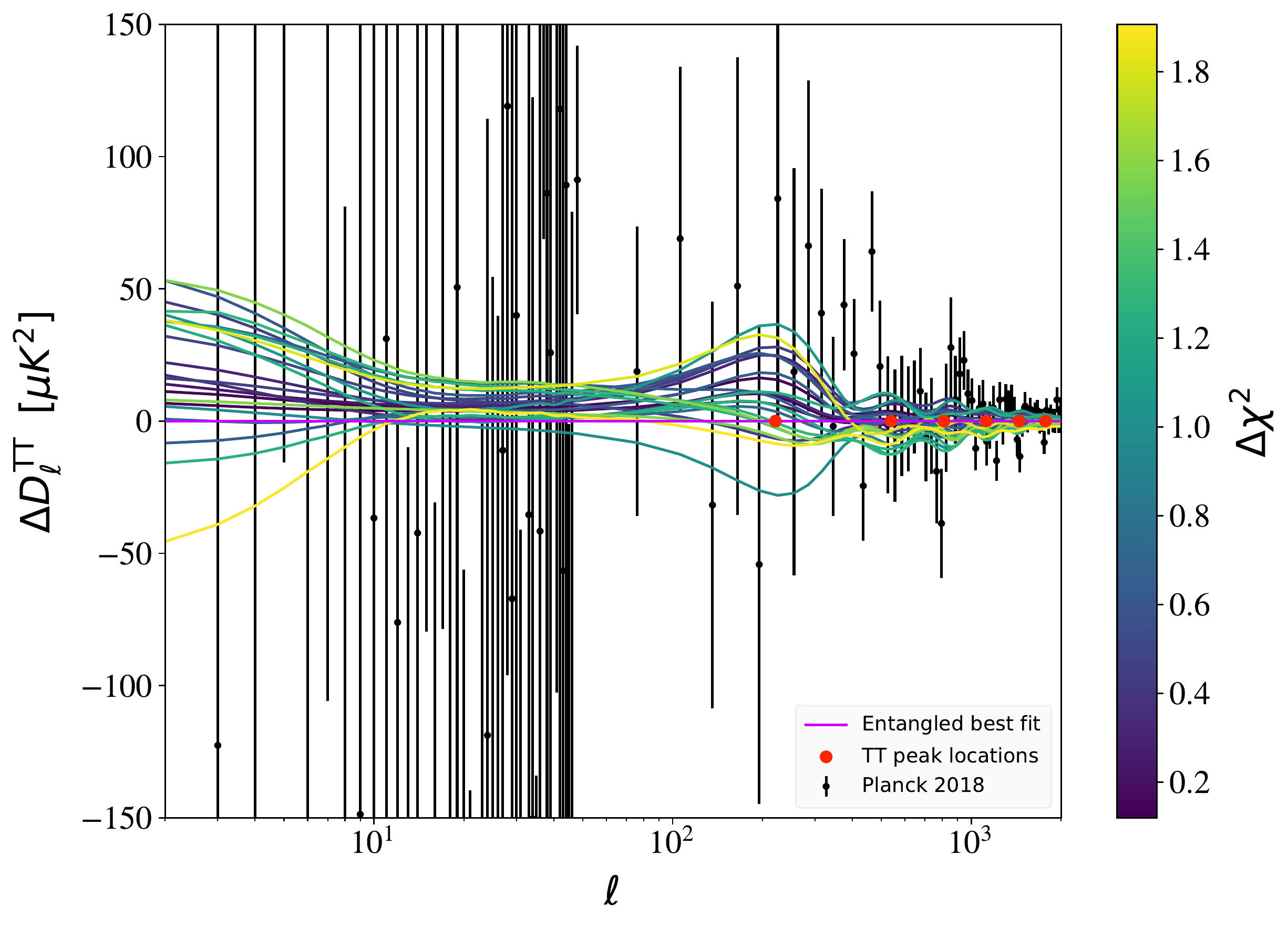} \hfill
\includegraphics[width=0.62\linewidth]{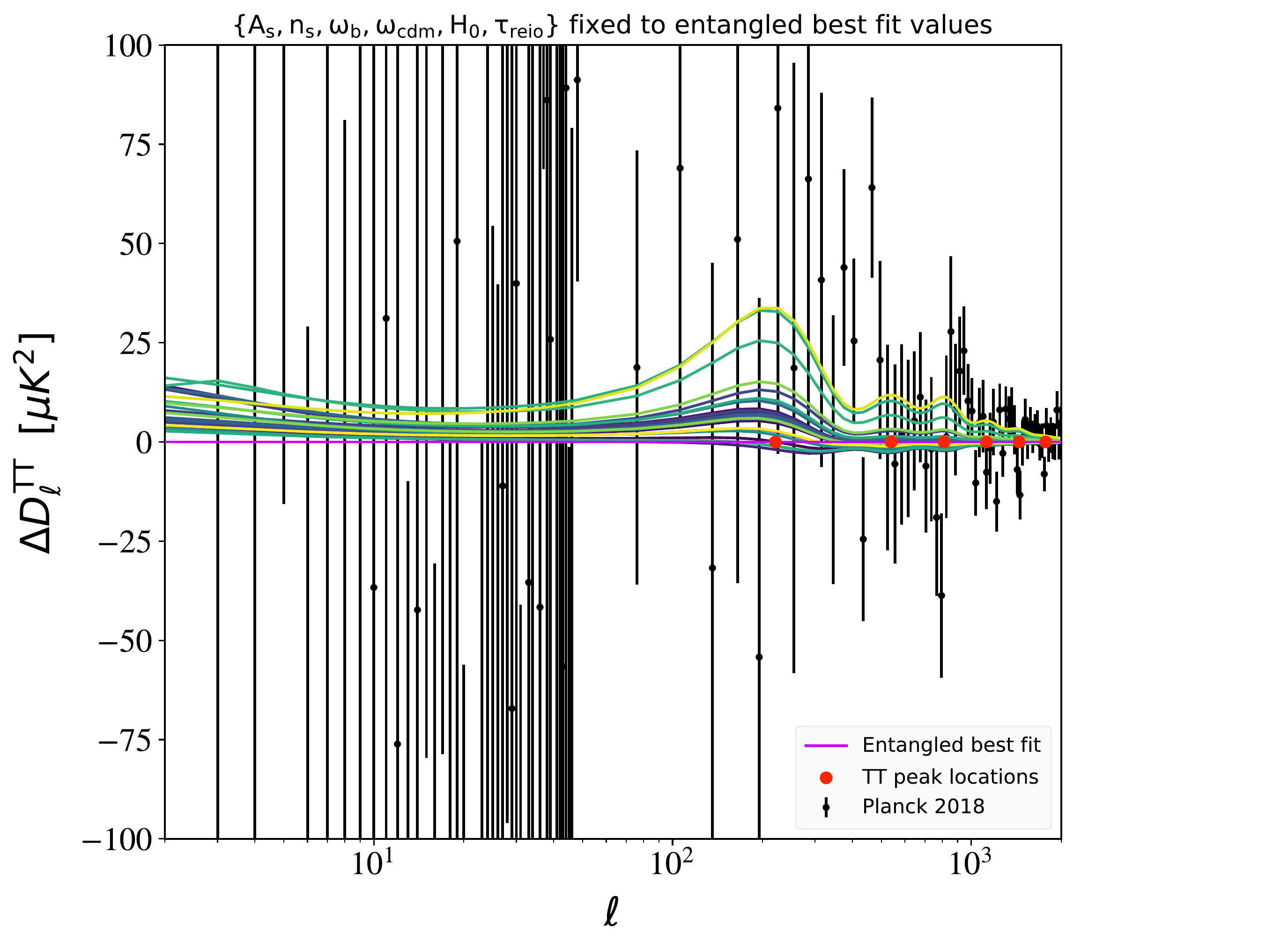}
\vspace{-0.15cm}
\caption{TT power spectrum (top) and residuals (middle) with respect to the entangled best fit value. The quantity $D_{\ell}^{TT} = \frac{\ell(\ell + 1)}{2 \pi} C_{\ell}^{TT}$. We plot $D_{\ell}^{TT}$ for the entangled best fit parameters, along with those parameters for which $\sup[\Delta_{s, {\rm norm}}^2] > 1.02$ and $\Delta \chi^{2} < 2$ (whose primordial spectra are shown in Fig~\ref{fig:k_ent_limits}). The locations of the peaks in the TT-spectra are plotted along with the residuals to guide the eye. The bottom plot investigates the effect of just the entangled parameters on the TT-spectra, as discussed in the text. Data is from the Planck 2018 data release.}
\label{fig:Cl_TT}
\end{figure}

We close this section by returning to the power spectra that show the largest deviations from BD within our perturbative approach. To explicitly demonstrate that even the highly deviating features of figure~\ref{fig:k_ent_limits} can masquerade in the CMB we plot the corresponding temperature $C_\ell$ spectrum in figure~\ref{fig:Cl_TT}. Clearly, the largest spread, by far, occurs in the low-$\ell$ regime which is precisely where the data has the largest sampling (``cosmic'') variance. Additionally, all the characteristic oscillatory features of the entangled power spectra have been washed out. This is largely due to the smearing effect of the window function, since the transfer function in eq.~\eqref{eq:cls} is given by
\vspace{-0.2cm}
\begin{equation}
    T_\ell (k, \overline{\eta_0}) = \int_{\overline{\eta}_{\rm initial}}^{\overline{\eta_0}} d \overline{\eta} S(k,\overline{\eta}) j_\ell[k(\overline{\eta_0} - \overline{\eta})]  
\end{equation}
where $S(k,\overline{\eta})$ contains all the cosmological hydrodynamics while the spherical Bessel function $j_\ell[k(\overline{\eta_0} - \overline{\eta})]$ acts as a window function encoding the geometric effect of projecting onto the CMB surface \cite{hu2002cosmic}. Since $j_\ell[..]$ has a non-zero width, it causes a transfer of power from a feature at a particular $k$-value to a range of $\ell$-values (in contrast to the $\delta$-function approximation where all the power is concentrated to a distinctive $\ell = k(\overline{\eta_0} - \overline{\eta}$)). In the same spirit, any oscillatory features in the primordial spectrum with frequency much less than the width of the window function see witness suppression \cite{chluba2015features}. A more thorough understanding regarding the kind of entanglement features that can survive the geometric effect of the transfer function is deferred to a future study. For this work, it suffices to mention that the apparent oscillations in the residuals shown in figure~\ref{fig:Cl_TT} should not be mistaken for the characteristic oscillations that appear in the primordial power spectrum.\footnote{Further isolating the effects of the entangled parameters $\mu$, $s_0$, and $k_{\mathrm{ent}}$ by fixing the rest of the parameters to their best fit values, as shown in the bottom part of figure~\ref{fig:Cl_TT}, supports this conclusion. Note that there is no color bar in the bottom plot of figure~\ref{fig:Cl_TT} since the $\chi^2$ was derived while varying all the cosmological parameters.} In fact, at least some of the variation in the residual temperature spectrum (middle panel in figure~\ref{fig:Cl_TT}) can be attributed to slight variations in the best-fit cosmological parameters, primarily in the $A_s - \tau_{\rm reio}$ plane, corresponding to a particular entangled primordial spectrum. 

\begin{table}[h]

	\heavyrulewidth=.08em
	\lightrulewidth=.05em
	\cmidrulewidth=.03em
	\belowrulesep=.65ex
	\belowbottomsep=0pt
	\aboverulesep=.4ex
	\abovetopsep=0pt
	\cmidrulesep=\doublerulesep
	\cmidrulekern=.5em
	\defaultaddspace=.5em
	\renewcommand{\arraystretch}{1.6}

	\begin{center}
		\small
		\begin{tabular}{lll}

			\toprule
		
			 {} & $\Lambda$CDM & Entangled \\
			\midrule
		
			\rowcolor[gray]{0.9}
				 Plik & 584.8 & 584.5
				\\[5mm]

				lowTT & 23.51 &  23.38
				\\[5mm]

			\rowcolor[gray]{0.9}
				lowEE & 395.9 & 396.3 
                \\[5mm]

				Total & 1004.2 & 1004.2 \\

 			\bottomrule
	
		\end{tabular}
	\end{center}
 \vspace{-0.3cm}
	\caption{\label{tab:lambdaCDMcompare} $\chi^{2}$ comparison for $\Lambda$CDM versus our entangled best fit parameters.}
	\end{table}

 \begin{table}[h]

	\heavyrulewidth=.08em
	\lightrulewidth=.05em
	\cmidrulewidth=.03em
	\belowrulesep=.65ex
	\belowbottomsep=0pt
	\aboverulesep=.4ex
	\abovetopsep=0pt
	\cmidrulesep=\doublerulesep
	\cmidrulekern=.5em
	\defaultaddspace=.5em
	\renewcommand{\arraystretch}{1.6}

	\begin{center}
		\small
		\begin{tabular}{llll}

			\toprule
		
			 Parameter & Posterior & Likelihood & Best Fit \\
			\midrule
		
			\rowcolor[gray]{0.9}
				 $10^{9} \mathrm{A}_{s}$ & $2.103^{+0.030}_{-0.034} \left( {}^{+0.070}_{-0.062}\right)$ & $\mathrm{---}$ & 2.106
				\\[5mm]

				$\mathrm{n}_{s}$ & $0.9653 \pm 0.0043(\pm 0.0084)$ &  $\mathrm{---}$ & 0.9653
				\\[5mm]

			\rowcolor[gray]{0.9}
				$\mathrm{log}_{10}(\mu)$ & $< -1.40(<-0.32)$ & U[-3,0] & -0.3002
                \\[5mm]

				$\mathrm{log}_{10}(s_{0})$ & $<-2.51(<0.10)$ & U[-6,2] & -5.482
                \\[5mm]

                 \rowcolor[gray]{0.9}
				$\mathrm{log}_{10}(k_{\mathrm{ent}})$ & U[-6,-2] & U[-6,-2] & -2.097
                \\[5mm]

				$\mathrm{log}_{10}(\lambda_{2,\mathrm{max}})$ & $>-4.33(>-6.83)$ & U[-8.6,-0.7] & -2.733 \\

 			\bottomrule
	
		\end{tabular}
	\end{center}
 \vspace{-0.4cm} 
	\caption{\label{tab:posteriorEtc} Summary statistics characterizing the distributions of various parameters discussed in section~\ref{sec:analysis}. For the near-Gaussian posteriors of $A_s$ and $n_s$, we quote the mean $\pm$ $68\% ($95\%$)$  central credible intervals. For the $\log{\mu}$, $\log{s_0}$, and $\log{\lambda_{2,\rm max}}$ ($\log{\lambda_{2,\rm max}}$ being a derived parameter) posteriors, we quote the $68\%$ ($95\%$) highest posterior density interval. Here, $U[..]$ denotes an approximately uniform distribution. }
	\end{table}

\vspace{-0.5cm} 
\section{Conclusions}

\label{sec:conclusions}

We have investigated the effects of entanglement in the inflationary universe on CMB observables. Even if inflation is mostly driven by a single light scalar field, many theories predict the existence of multiple, heavier degrees of freedom. As demonstrated in ref.~\cite{Baunach:2021yvu}, if there exist other spectator fields, an entangled state is expected to emerge, which could in principle induce some observable imprints in the CMB. Also, as discussed in the introduction, from the point of view of constructing the most general Gaussian state respecting all the symmetries of the system, we need to include the entanglement kernel. 

Here we have focused on the simple yet non-trivial situation in which the spectator field starts away from its minimum, but with zero initial velocity. To enable tractable calculations we worked within a perturbative framework.  As demonstrated in figure~\ref{fig:Muvariation}, deviations from the purely Bunch--Davies predictions for the primordial power spectrum are expected in the form of a modulated oscillatory signal. By performing a Monte Carlo parameter estimation analysis, we further probed the effects of such oscillations on the CMB observables and found that within our framework our predictions are largely compatible with the Planck data. 

One immediate consequence is that low levels of entanglement can actively masquerade in the CMB, in such a way that we may be unable to distinguish it from purely single-field inflation models from power spectrum observables alone. In particular, we have found that the kinds of features introduced in the primordial power spectrum within our framework are unable to confuse the inference of the usual six standard cosmology parameters. However, remarkably, even when the cosmological parameters are held approximately fixed, our model allows for significant changes to the primordial spectrum, with negligible changes to the Planck $\chi^2$ budget. In other words, current constraints are unable to rule out entanglement, even for the simplest dynamics that we have considered in this paper.

Based on these results, one natural direction to follow up would be to look for other probes to disentangle this effect, which could include the bispectrum of perturbations or the matter power spectrum. Another interesting direction would be to examine the effects of going beyond just a free massive scalar as a spectator, but perhaps looking at an axionic-type potential, as well as entanglement with other spin fields. This latter approach might yield an interesting line of sight into some of the large scale anomalies in CMB data. It could also be interesting to extend this work by developing a complimentary technical set-up in Heisenberg picture, which would allow one to explore concepts of entanglement in that formalism. Such an investigation might reveal 
underlying entanglement in other work investigating the imprints of spectator fields during inflation, even though the authors had not previously considered their results from that conceptual lens. Moreover, increased numerical efficiency and/or processing power could allow us to expand the limits of our framework in directions that would allow the data to be more informative. To reiterate our main point, CMB data alone are consistent with the BD initial state, but \emph{cannot} rule out entangled states, even ones with significant amounts of entanglement as measured by the primordial power spectrum. 

\acknowledgments
We thank Lloyd Knox and Marius Millea for useful discussions.  This work was supported in part by the
U.S. Department of Energy, Office of Science, Office of
High Energy Physics QuantISED program under Contract No. KA2401032. Some of our computations were performed on the UC Davis Peloton computer cluster on a node purchased with funds from DOE Office of Science award DE-SC0009999.

\appendix

\section*{Supplementary Results}

Here we collect additional results that the main text makes reference to. We also summarize some additional technical extensions to our work that, while not used for the parameter estimation analysis, offer a complement to the reader interested in the details of our entangled states formalism.

\section{Free massive scalar zero mode analytic solution}

\label{app:free_massive}

The equation that describes the classical evolution of the zero mode of a massive spectator field in an expanding spacetime---in terms of the dimensionless parameters described in section~\ref{sec:kerneldimless}---is given by:
\begin{equation}
    \label{eq:dimlessBackgroundA}
    s''(\tau) - \frac{2}{\tau(1-\epsilon)} s'(\tau) + \frac{\mu^{2} \partial_{s}V(s)}{\tau^{2}(1-\epsilon)^{2}} = 0 \ .
\end{equation}
For the free massive scalar potential, given by $V(s) = \frac{1}{2} s^{2}$ in dimensionless quantities, this has an analytic solution.  Given initial conditions $s(\tau_0 = -1) = s_0$, $\partial_{\tau}s(\tau_0 = -1) = v_0$, the solutions are:
\begin{subequations}
\label{eq:analyticSspr}
\begin{align}
s(\tau) & = s_{+}(-\tau)^{p_{+}} + s_{-}(-\tau)^{p_{-}} \\
s'(\tau) & = - s_{+}p_{+}(-\tau)^{p_{+}-1} - s_{-}p_{-}(-\tau)^{p_{-}-1}
\end{align}
\end{subequations}
with
\begin{equation}
\label{eq:Ppm}
p_{\pm} = \frac{(3 - \epsilon) \pm \sqrt{(3 - \epsilon)^{2} - 4 \mu^{2}}}{2(1-\epsilon)}
\end{equation}
and
\begin{subequations}
\label{eq:Spm}
\begin{align}
s_{+} & = \frac{p_{-}s_0 + v_0}{p_{-} - p_{+}} \\
s_{-} & = \frac{-p_{+}s_0 - v_0}{p_{-} - p_{+}} \ .
\end{align}
\end{subequations}
Having this analytic solution gives some computation speedup in calculating the entangled power spectrum.  It also enables the analytic treatment discussed in appendix~\ref{app:analytic}---which is only possible if eq.~\eqref{eq:dimlessBackgroundA} can be solved analytically. 

Additionally, one can use eq.~\eqref{eq:analyticSspr} to exactly specify the $\lambda$ parameters for a free massive scalar field:
\begin{subequations}
\label{eq:dimlessLambdaFMS}
\begin{align}
 \lambda_1 & = \frac{(1-\epsilon)}{\sqrt{2 \epsilon}} (-\tau) \partial_{\tau}s = \frac{(1-\epsilon)}{\sqrt{2 \epsilon}} \left[- s_{+}p_{+}(-\tau)^{p_{+}} - s_{-}p_{-}(-\tau)^{p_{-}}\right]
\\
\lambda_2 & =  \frac{\mu^{2}}{\sqrt{2 \epsilon}} \ \partial_{s}V(s) = \frac{\mu^{2}}{\sqrt{2 \epsilon}} \left[s_{+}(-\tau)^{p_{+}} + s_{-}(-\tau)^{p_{-}} \right] \ .
\end{align}
\end{subequations}
For cases when $v_0 = 0$---which we explore in this paper's Monte Carlo analysis---one can derive the maximum value of the $\lambda$ parameters during the course of inflation:
\begin{subequations}
\label{eq:lambdaMax}
\begin{align}
\lambda_{1,max} & =  \left| \frac{(1-\epsilon)}{\sqrt{2 \epsilon}} s_0 \frac{p_{+}p_{-}}{p_{-} - p_{+}} \left[ -\left( \frac{p_{+}}{p_{-}} \right)^{\frac{p_{+}}{p_{-} - p_{+}}} + \left( \frac{p_{+}}{p_{-}}\right)^{\frac{p_{-}}{p_{-} - p_{+}}} \right] \right| \\
\lambda_{2,max} & =  \frac{\mu^{2} s_0}{\sqrt{2 \epsilon}} 
\end{align}
\end{subequations}
by finding the time at which $\frac{\partial \lambda_{1,2}}{\partial \tau} = 0$ and then evaluating eq.~\eqref{eq:dimlessLambdaFMS}. Furthermore, one can verify that $\lambda_{2,max} < 1$ will always be the more restrictive condition for a given $\mu$, given $v_0 = 0$.

\section{Entanglement kernel analytic solutions and super-Hubble scale spectator masses}

\label{app:analytic}

In this appendix we discuss two things---the possibility of analytic solutions for the $\lambda$ expanded power spectrum for a free massive scalar spectator field, and what these analytic solutions might tell us about spectator masses with $\mu = \frac{m_{\sigma}}{H_{ds}} > 1$.

First, note that for a free massive scalar spectator with $V(s) = \frac{1}{2} s^2$, both eqs.~\eqref{eq:dimlessBackground} and \eqref{eq:dimlessC1} admit analytic solutions.  Solutions for the spectator zero mode are discussed in appendix~\ref{app:free_massive}, here we discuss the exact solutions to $C_{q}^{(1)}(\tau)$, which are only possible if eq.~\eqref{eq:dimlessBackground} can be solved analytically.

The easiest thing to do is to solve eq.~\eqref{eq:dimlessC1} via an integrating factor solution.  If one rewrites that equation as:
\begin{align}
\label{eq:C1intfac}
\partial_{\tau} C_{q}^{(1)} + i C_{q}^{(1)}\left(A_{q}^{(0)} + B_{q}^{(0)} \right) & = 
 -i \tilde{\lambda}_1 A_{q}^{(0)}B_{q}^{(0)}  + \frac{(A_{q}^{(0)} - B_{q}^{(0)} )}{2(1-\epsilon)\tau} \left[ \left(3-\epsilon+\frac{\eta_{sl}}{2}\right) \tilde{\lambda}_1 + \tilde{\lambda}_2 \right] \notag \\
& {} {} \quad  -i \tilde{\lambda}_1 \left[ \left(\frac{q}{1-\epsilon}\right)^2 +\frac{\mu^{2} \partial_{s}^{2}V(s)}{2 (1-\epsilon)^{2} \tau^{2}} +\frac{1 +\frac{5}{4}\eta_{sl}}{(1-\epsilon)^{2}\tau^{2}} \right] \notag \\
& {} {} \quad -i \tilde{\lambda}_2 \left[\frac{1 + \epsilon + \frac{\eta_{sl}}{2}}{(1-\epsilon)^{2}\tau^{2}} \right]
\end{align}
one can see it of the form
\begin{equation}
\label{eq:IntfacEq}
\partial_{\tau} C_{q}^{(1)} + P(\tau)C_{q}^{(1)} = Q(\tau)
\end{equation}
which has a solution given by:
\begin{equation}
\label{eq:IntfacSol}
C_{q}^{(1)}(\tau) = e^{-\int P(\tau) d\tau} \left( \int Q(\tau)  e^{\int P(\tau) d\tau} d\tau \right) + C  e^{-\int P(\tau) d\tau} \ 
\end{equation}
and $C$ without subscript is a constant of integration.

If one expresses the dimensionless zeroth order kernels as:
\begin{subequations}
\label{eq:A0B0expand}
\begin{align}
A_{q}^{(0)}(\tau) & = -i\frac{f_{v}'(\tau)}{f_{v}(\tau)} = i\left[ \frac{(1 - 2 \nu_{f})}{2(-\tau)} + \frac{q}{(1-\epsilon)} \frac{H^{(2)}_{\nu_{f}-1} \left(\frac{-q \tau}{(1-\epsilon)}\right)}{H^{(2)}_{\nu_{f}}\left(\frac{-q \tau}{(1-\epsilon)}\right)} \right] \\
B_{q}^{(0)}(\tau) & = -i\frac{g_{\theta}'(\tau)}{g_{\theta}(\tau)} = i\left[ \frac{(1 - 2 \nu_{g})}{2(-\tau)} + \frac{q}{(1-\epsilon)} \frac{H^{(2)}_{\nu_{g}-1} \left(\frac{-q \tau}{(1-\epsilon)}\right)}{H^{(2)}_{\nu_{g}}\left(\frac{-q \tau}{(1-\epsilon)}\right)} \right]
\end{align}
\end{subequations}
with $\nu_f$ and $\nu_g$ given in eqs.~\eqref{eq:Hankelnuf} and \eqref{eq:Hankelnug}, deriving the integrating factor proceeds straightforwardly. The result is:
\begin{equation}
\label{eq:Intfac}
e^{\int P(\tau) d\tau} = (-\tau) \left(\frac{q}{1 - \epsilon}\right)^{\nu_f + \nu_g} \left[H^{(2)}_{\nu_{f}}\left(\frac{-q \tau}{(1-\epsilon)}\right) H^{(2)}_{\nu_{g}}\left(\frac{-q \tau}{(1-\epsilon)}\right) \right] \ .
\end{equation}
After some fairly intensive algebra, one finds the full solution is given by:
\begin{equation}
\label{eq:C1analytic}
C_{q}^{(1)}(\tau)  = \left(\frac{q}{1 - \epsilon}\right)^{-\nu_f - \nu_g} \frac{\left[T(\tau) - T(\tau_0 = -1)\right]}{(-\tau)\left[H^{(2)}_{\nu_{f}}\left(\frac{-q \tau}{(1-\epsilon)}\right) H^{(2)}_{\nu_{g}}\left(\frac{-q \tau}{(1-\epsilon)}\right) \right]}
\end{equation}
where the constant of integration was determined by the initial condition $C_{q}^{(1)}(\tau_0 = -1) = 0$ and $T(\tau)$ is given by:
\begin{align}
\label{eq:Ttau}
T(\tau) & = \frac{i}{\lambda \sqrt{2 \epsilon}}  \left(\frac{q}{1 - \epsilon}\right)^{\nu_f + \nu_g} \Bigg[ \mathcal{I}_{(p_{+}-1,\nu_f, \nu_g )} s_{+} \left(\frac{1-\epsilon}{q}\right)^{p_{+}} \Bigg( \frac{(1-\epsilon)(1-2\nu_f)(1-2\nu_g)}{4}p_{+} \notag \\
& {} {} \quad - \frac{\nu_g - \nu_f}{2(1-\epsilon)} \left[ \left(3 - \epsilon + \frac{\eta}{2}\right)(1-\epsilon)p_{+} - \mu^{2} \right] - \frac{(1-\epsilon)\left(1 + \frac{5 \eta_{sl}}{4} + \frac{\mu^{2}}{2} \right) p_+ - \left(1 + \epsilon + \frac{\eta}{2} \right)\mu^{2}}{(1-\epsilon)^{2}} \Bigg) \notag \\
{} {} \quad & + \mathcal{I}_{(p_{-}-1,\nu_f, \nu_g )} s_{-} \left(\frac{1-\epsilon}{q}\right)^{p_{-}} \Bigg( \frac{(1-\epsilon)(1-2\nu_f)(1-2\nu_g)}{4}p_{-} \notag \\
& {} {} \quad - \frac{\nu_g - \nu_f}{2(1-\epsilon)} \left[ \left(3 - \epsilon + \frac{\eta}{2}\right)(1-\epsilon)p_{-} - \mu^{2} \right] - \frac{(1-\epsilon)\left(1 + \frac{5 \eta_{sl}}{4} + \frac{\mu^{2}}{2} \right) p_{-} - \left(1 + \epsilon + \frac{\eta}{2} \right)\mu^{2}}{(1-\epsilon)^{2}} \Bigg) \notag \\ 
{} {} \quad & - \frac{(1 - \epsilon)^{p_+ + 1}}{q^{P_+}} (s_+ p_+) \mathcal{I}_{(p_{+}+1,\nu_f, \nu_g )} - \frac{(1 - \epsilon)^{p_- + 1}}{q^{P_-}} (s_- p_-) \mathcal{I}_{(p_{-}+1,\nu_f, \nu_g )} \notag \\
{} {} \quad & + \mathcal{I}_{(p_{+},\nu_f, \nu_{g} - 1 )} s_{+}\left(\frac{1-\epsilon}{q}\right)^{p_{+}} \Bigg( \frac{(1-\epsilon)(1-2 \nu_f)}{2} p_{+} + \frac{ \left[ \left(3 - \epsilon + \frac{\eta_{sl}}{2} \right)(1-\epsilon)p_{+} - \mu^2 \right] }{2(1-\epsilon)} \Bigg) \notag \\
{} {} \quad & + \mathcal{I}_{(p_{-},\nu_f, \nu_{g} - 1 )} s_{-}\left(\frac{1-\epsilon}{q}\right)^{p_{-}} \Bigg( \frac{(1-\epsilon)(1-2 \nu_f)}{2} p_{-} + \frac{ \left[ \left(3 - \epsilon + \frac{\eta_{sl}}{2} \right)(1-\epsilon)p_{-} - \mu^2 \right] }{2(1-\epsilon)} \Bigg) \notag \\
{} {} \quad & + \mathcal{I}_{(p_{+},\nu_g, \nu_{f} - 1 )} s_{+}\left(\frac{1-\epsilon}{q}\right)^{p_{+}} \Bigg( \frac{(1-\epsilon)(1-2 \nu_g)}{2} p_{+} - \frac{ \left[ \left(3 - \epsilon + \frac{\eta_{sl}}{2} \right)(1-\epsilon)p_{+} - \mu^2 \right] }{2(1-\epsilon)} \Bigg) \notag \\
{} {} \quad & + \mathcal{I}_{(p_{-},\nu_g, \nu_{f} - 1 )} s_{-}\left(\frac{1-\epsilon}{q}\right)^{p_{-}} \Bigg( \frac{(1-\epsilon)(1-2 \nu_g)}{2} p_{-} - \frac{ \left[ \left(3 - \epsilon + \frac{\eta_{sl}}{2} \right)(1-\epsilon)p_{-} - \mu^2 \right] }{2(1-\epsilon)} \Bigg) \notag \\
{} {} \quad & + (1-\epsilon)s_{+}p_{+} \left(\frac{1-\epsilon}{q}\right)^{p_{+}} \mathcal{I}_{(p_{+}+1,\nu_{f} - 1, \nu_{g} - 1 )} + (1-\epsilon)s_{-}p_{-} \left(\frac{1-\epsilon}{q}\right)^{p_{-}} \mathcal{I}_{(p_{-}+1,\nu_{f} - 1, \nu_{g} - 1 )} \Bigg] \ .
\end{align}
The parameters $s_{\pm}$ and $p_{\pm}$ are defined in appendix~\ref{app:free_massive}, $\lambda$ is our expansion parameter (as discussed in section~\ref{sec:theory}), and $\mathcal{I}_{(a,b,c)}$ is defined as:
\begin{equation}
\label{eq:Iint}
\mathcal{I}_{(a,b,c)} = \left . \left[\int x^{a} H_{b}^{(2)}(x) H_{c}^{(2)}(x) dx \right] \right|_ {x=\frac{-q \tau}{(1-\epsilon)}}
\end{equation}
which can be exactly evaluated---using Mathematica or other methods of your choice---to be a combination of power laws, generalized hypergeometric functions and gamma functions. The exact evaluated form of $\mathcal{I}_{(a,b,c)}$ is not particularly illuminating, but what is interesting is that eq.~\eqref{eq:Iint} together with eqs.~\eqref{eq:C1analytic} and \eqref{eq:Ttau} hint that much of the oscillatory nature of our entanglement kernel $C_{q}^{(1)}$---both in time and as it contributes to $\Delta_{s}^{2}(k)$---is sourced by integrals that mix the two Hankel functions from the inflaton and spectator fields.

One might hope to gain further intuition and derive an analytic expression for the power spectrum in eq.~\eqref{eq:dimlessPS2}  by also solving the $A_{q}^{(2)}$ equation in the same way, however this is only partially possible.  The contribution to $A_{q}^{(2)}$ from the $C$ kernel independent terms in eq.~\eqref{eq:dimlessA2} can also be solved analytically, using similar methods to what was described above for $C_{q}^{(1)}$. However, the contribution from the C kernel dependent terms in that equation cannot be solved for analytically---as far as the authors know at this time of writing---due several integrals over multiple generalized hypergeometric functions that show up due to eq.~\eqref{eq:Iint}.

\begin{figure}[!h]
\centering
\includegraphics[width=0.67\linewidth]{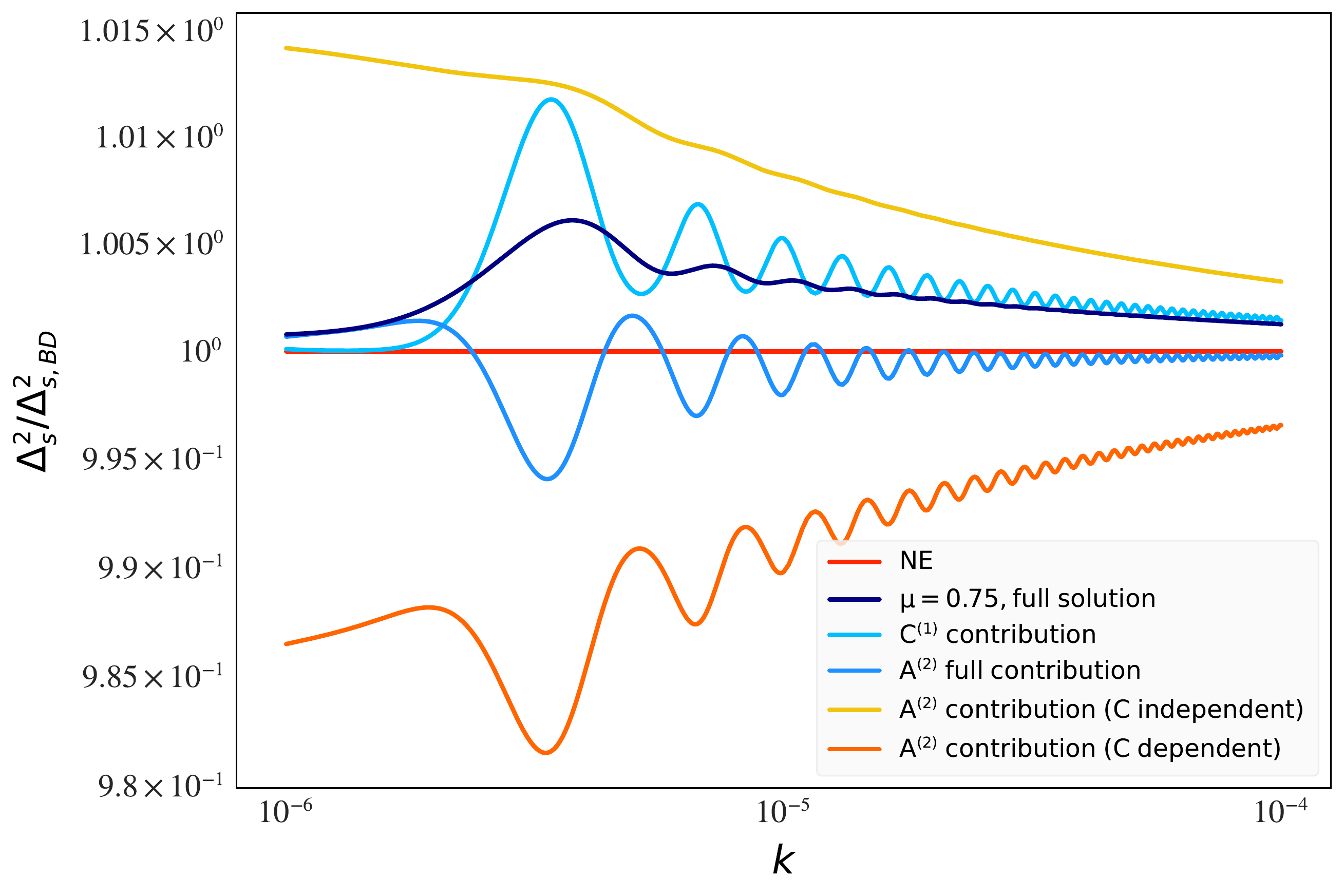}
\caption{Log-log plot of $\Delta_{s, {\rm norm}}^2$ for $\mu = 0.75$, $s_0 = \frac{0.2\sqrt{2 \epsilon}}{\mu^{2}}$ (with $\epsilon = 10^{-7}$) and $v_0 =0$. We plot the full solution for $\Delta_{s, {\rm norm}}^2$ along with its `component parts,' as discussed in the text. The non entangled case corresponds to $\Delta_{s, {\rm norm}}^2= 1$, and we take our expansion parameter to be $\lambda = \lambda_{2,\mathrm{max}}$, where $\lambda_{2,\mathrm{max}}$ is defined in eq.~\eqref{eq:lambdaMax}.}
\label{fig:PsParts}
\end{figure}

But even though a full analytic treatment of $\Delta_{s}^{2}(k)$ is not currently possible, it turns out that the C kernel analytic solution derived here can give a baseline estimate of what the overall behavior of the power spectrum will be.  Consider the equation for $\Delta_{s, {\rm norm}}^2$, i.e.
\begin{equation}
\label{eq:dimlessPS2A}
\Delta_{s, {\rm norm}}^2 = \left[1 + \lambda^{2} \left(\frac{-A_{k R}^{(2)}}{A_{k R}^{(0)}} +\frac{(C_{k R}^{(1)})^2}{A_{k R}^{(0)} B_{k R}^{(0)}} \right) \right] =\left[1 + \lambda^{2} \left(\frac{-A_{q R}^{(2)}}{A_{q R}^{(0)}} +\frac{(C_{q R}^{(1)})^2}{A_{q R}^{(0)} B_{q R}^{(0)}} \right) \right] 
\end{equation}
which is just a rewritten form of eq.~\eqref{eq:dimlessPS2}.
Figure~\ref{fig:PsParts} plots the full numerical solution to $\Delta_{s, {\rm norm}}^2$, along with its `component parts'
---the contribution from just $C_{q}^{(1)}$ alone (which can be solved numerically or analytically), the contribution from the C kernel independent part of $A_{q}^{(2)}$ (which can also be solved by either method), and the C dependent part of $A_{q}^{(2)}$ (which can only be evaluated numerically). Even though the full solution to $\Delta_{s, {\rm norm}}^2$ is different than that given by the $C^{(1)}$ contribution---$\left[1 + \lambda^{2} \left(\frac{(C_{q R}^{(1)})^2}{A_{q R}^{(0)} B_{q R}^{(0)}} \right) \right] $---the latter still hints at many of the features the full solution contains, namely oscillations in k and a decaying exponential envelope for this choice of $\mu$, $s_0$ and $v_0$.

One can then use the C kernel analytic solutions to estimate the effects of entanglement on the power spectrum for spectators with $\mu = \frac{m_{\sigma}}{H_{ds}} > 1$. These solutions are potentially interesting, but not easily amenable to numerical evaluation due to the fact that the Hankel function index for the spectator mode, $\nu_g$, quickly becomes imaginary for $\mu > 1$. Hankel functions of imaginary order can be evaluated in Mathematica, but we have not found a way to evaluate these functions sufficiently rapidly to use in our Monte Carlo calculations.

\begin{figure}[!h]
\centering
\includegraphics[width=0.8\linewidth]{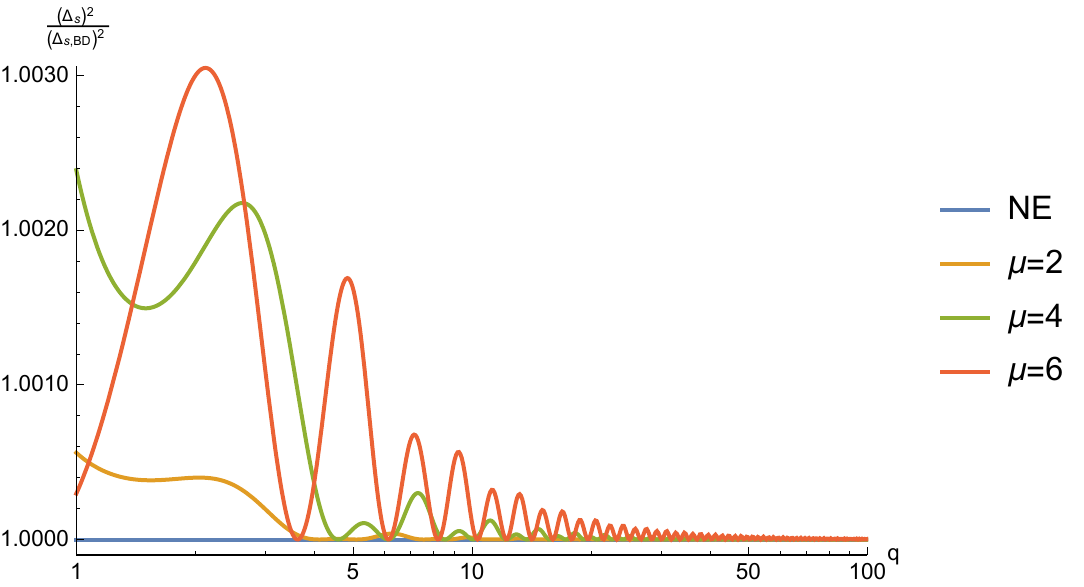}
\caption{Log-log plot of the C kernel contribution to $\Delta_{s, {\rm norm}}^2$, $\left[1 + \lambda^{2} \left(\frac{(C_{q R}^{(1)})^2}{A_{q R}^{(0)} B_{q R}^{(0)}} \right) \right] $, for $\mu = 2$,4,6. We take $s_0 = \frac{0.2\sqrt{2 \epsilon}}{4^{2}}$ (with $\epsilon = 10^{-7}$) and $v_0 =0$ for all three curves, to enable easier comparison. $q = k \times 10^{6}$ on the horizontal axis, and the non entangled case corresponds to $\Delta_{s, {\rm norm}}^2 = 1$. (We take our expansion parameter to be $\lambda = \lambda_{2,\mathrm{max}}$, where $\lambda_{2,\mathrm{max}}$ is defined in eq.~\eqref{eq:lambdaMax}. )}
\label{fig:SuperHubble}
\end{figure}

So, as a preview of possible extensions to our current work, in figure~\ref{fig:SuperHubble} we plot the C kernel analytic contribution to the power spectrum, $\left[1 + \lambda^{2} \left(\frac{(C_{q R}^{(1)})^2}{A_{q R}^{(0)} B_{q R}^{(0)}} \right) \right] $, for $\mu = 2$, 4 and 6. This plot shows an example of how different the entanglement structure may be for super Hubble scale masses.

\section{Prior volume weighting}
\label{sec:appendix-prior-volume}

In this section we further elucidate the evidence for prior effects, in particular the effect of prior volume weighting, that was alluded to in section~\ref{sec:analysis}. Recall that the posteriors are prior driven due to the effects of imposing two conditions: that $\mu<1$ and that the spectator be subdominant to the inflaton energy so that, for a given $\mu$, $s0< 0.5\sqrt{2\epsilon}/\mu^2$---which is effectively a joint prior on $\mu$ and $s_0$. Meanwhile, there are also the independent uniform priors (the region from which the Monte Carlo will draw samples): $\log{\mu} \in [-3,0]$ and $\log{s_0} \in [-6,2]$ where the lower bounds stem from our motivation to exclude regions of very small mass that are practically degenerate with the BD state (i.e. deviate negligibly). However, this choice of limiting the lower bound in $s_0$ ($\mu$) results in less prior weight to the larger values of $\mu$ ($s_0$). To illustrate this point, suppose that we instead sample $\log{s_0} \in [-20,2]$ so that the higher $\mu$ values now have more prior volume available than before. The resulting posterior distribution (again, assuming a uniform likelihood on all the parameters) shows an increased probability density towards larger values of $\mu$; in fact, the posterior is now much closer to a uniform distribution. 

Note that often a stricter prior is chosen precisely to mitigate the effect of prior volume effects (see e.g. section III in \cite{smith2021-EDE}). In such cases, expanding the prior range will further exacerbate the issue. However, in our case, not only is the data uninformative on the parameters across the range of variation, the joint prior complicates matters by assigning less prior volume to the higher $\mu$ regime so that extending the lower bound in the independent $\log{s_0}$ prior can provide some compensation to alleviate the effect of the joint prior. Of course, because of the symmetry between $s_0$ and $\mu^2$ in the joint prior, one can achieve the same effect on the marginalized $\log{s_0}$ posterior by extending the lower bound of the independent $\log{\mu}$ prior. 

\begin{figure}[!h]
\centering
\includegraphics[width=0.75\linewidth]{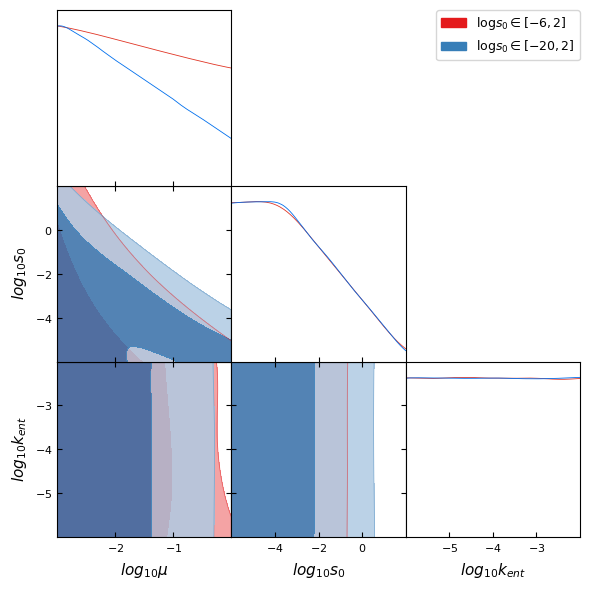}
\caption{The effect of decreasing the lower bound of $\log{s_0}$ can be clearly seen in the marginalized $\log{\mu}$ posterior.}
\label{fig:only_prior_extended_s0}
\end{figure}

\newpage

\bibliographystyle{unsrt}
\bibliography{ZetaSpecEnt.bib}







\end{document}